\documentclass[10pt]{article}
\usepackage{amssymb}
\usepackage{}
\usepackage{amsfonts}
\usepackage{bbm}
\usepackage{mathrsfs}
\usepackage{mathrsfs}
\usepackage{longtable}
\usepackage{supertabular}
\usepackage{amsfonts,mathrsfs,latexsym,amsmath,amssymb,amsthm}

\begin{document}
\title{{\bf Quasi-Cyclic Codes Over Finite Chain Rings}}
\author{{\bf  Jian Gao, ~LinZhi Shen, ~Fang-Wei Fu }\\
 {\footnotesize \emph{Chern Institute of Mathematics and LPMC, Nankai University}}\\
  {\footnotesize  \emph{Tianjin, 300071, P. R. China}}\\
  {\footnotesize  \emph{ Email:~jiangao@mail.nankai.edu.cn}}}
\date{}

\maketitle \noindent {\small {\bf Abstract} In this paper, we mainly consider quasi-cyclic (QC) codes over
finite chain rings. We study module structures and trace representations of QC codes, which lead to some lower bounds on the minimum Hamming distance of QC codes. Moreover, we investigate the structural properties of $1$-generator QC codes. Under some conditions, we discuss the enumeration of $1$-generator QC codes and describe how to obtain the one and only one generator for each $1$-generator QC code.}
 \vskip 1mm



\vskip 3mm \baselineskip 0.2in

\vskip 3mm \noindent {\bf 1 Introduction} \vskip 3mm \noindent
A finite commutative ring with identity is called a \emph{finite chain ring} if its ideals are linearly ordered by
inclusion. It is well known that every ideal of finite chain ring is principal and its maximal ideal is unique. Let $R$ denote the finite chain ring, $\gamma$ a generator of its maximal ideal and $\mathbb{F}$ the residue field $R/\langle \gamma \rangle$. The ideals of $R$ form a chain as follows
$$ \langle 0 \rangle=\langle \gamma^s\rangle\subseteq\langle \gamma^{s-1}\rangle\subseteq \cdots \subseteq \langle \gamma \rangle \subseteq \langle 1 \rangle=R.$$ The integer $s$ is called the \emph{nilpotency index} of $R$. If $\mathbb{F} \cong \mathbb{F}_{q}$, then $|R|=q^s$. In this paper, we assume that $\mathbb{F}$ is equivalent to $\mathbb{F}_{q}$.

\vskip 3mm \par The classical examples of finite chain rings that are not finite fields are the integer residue ring $\mathbb{Z}_{p^s}$, the Galois ring ${\rm GR}(p^m, s)$ and the ring $\mathbb{F}_{p^m}+u\mathbb{F}_{p^m}+\cdots+u^{s-1}\mathbb{F}_{p^m}$, where $p$ is a prime number and $m$, $s$ are positive integers such that $s\geq 2$. Note that the ring $\mathbb{F}_{p^m}+u\mathbb{F}_{p^m}+\cdots+u^{s-1}\mathbb{F}_{p^m}$ is isomorphic to $\mathbb{F}_{p^m}[u]/\langle u^s\rangle$, the only finite chain ring with character $p$ and nilpotency index $s$. Define the ring epimorphism $^- : R\rightarrow \overline{R}=\mathbb{F}_q$ by $r\mapsto \overline{r}$, where $\overline{r}$ denotes $r+\langle \gamma\rangle$. Extending the ring epimorphism $^- : R[x]\rightarrow \mathbb{F}_q[x]$ by $r_0+r_1x+\cdots+r_nx^n\mapsto \overline{r}_0+\overline{r}_1x+\cdots+\overline{r}_nx^n$, and the image of $f(x)\in R[x]$ under the map $^-$ is denoted by $\overline{f}(x)\in \mathbb{F}_q[x]$.

\vskip 3mm \par
Let $f(x)$ and $g(x)$ be polynomials of $R[x]$. A monic polynomial $d(x)$ is called the \emph{greatest common divisor} of $f(x)$ and $g(x)$ if $d(x)$ is a divisor of $f(x)$ and $g(x)$, and if $e(x)$ is a divisor of $f(x)$ and $g(x)$, then $e(x)$ is a divisor of $d(x)$. We denote $d(x)={\rm gcd}(f(x), g(x))$. Two polynomials $f(x)$ and $g(x)$ are said to be \emph{coprime} over $R$ if there are two polynomials $a(x)$ and $b(x)$ in $R[x]$ such that $a(x)f(x)+b(x)g(x)=1$. It is well known that $f(x)$ and $g(x)$ are coprime over $R$ if and only if $\overline{f}(x)$ and $\overline{g}(x)$ are coprime over $\mathbb{F}_q$. An interesting thing is  that in $R[x]$ two coprime polynomials may have a common divisor with degree $\geq 1$. However, it is clear that the common divisor must be a unit in $R[x]$. Therefore if $f(x)$ and $g(x)$ are monic polynomials over $R$, then their common divisor is only $1$ for the case coprime of them.

\vskip 3mm \par
A polynomial $f(x)\in R[x]$ is said to be \emph{basic irreducible} if $\overline{f}(x)$ is irreducible in $\mathbb{F}_q[x]$, and \emph{basic primitive} if $\overline{f}(x)$ is primitive in $\mathbb{F}_q[x]$. If $f(x)$ is a monic basic irreducible polynomial with degree $m$ over $R$, then the residue class ring $R[x]/\langle f(x)\rangle$ is called the $m$-th \emph{Galois extension ring} of $R$, and denoted as ${\mathcal R}$. ${\mathcal R}$ is also a finite chain ring, with maximal ideal $\langle \gamma \rangle$ and nilpotency index $s$. If $\xi$ is a root of $f(x)$, then ${\mathcal R}=R[\xi]$, i.e., ${\mathcal R}$ is a \emph{free module} of rank $m$ over $R$ with $\{1, \xi, \ldots, \xi^{m-1}\}$ as a basis. If $f(x)$ is a basic primitive polynomial over $R$, and $\xi$ is the root of $f(x)$, then the order of $\xi$ is $q^m-1$. Let the \emph{Teichm\"{u}ller set} be ${\mathcal T}=\{0, 1,\xi, \ldots, \xi^{q^m-2} \}$. Then each element $r$ of ${\mathcal R}$ can be expressed uniquely as $$r=r_0+r_1\gamma+\cdots+r_{s-1}\gamma^{s-1},$$ where $r_0,r_1,\ldots,r_{s-1}\in {\mathcal T}$. The units of ${\mathcal R}$ form a multiplicative group denoted by ${\mathcal R}^*$, which is a direct product $\langle \xi \rangle \times G$, where $G=\{ 1+\pi | \pi \in \langle \gamma \rangle\}$ is a group of order $q^{(s-1)m}$.

\vskip 3mm \par

Let $R$ be a finite chain ring and $R^n$ be the set of $n$-tuples over $R$. $\mathscr{C}$ is a \emph{linear code} of length $n$  over $R$ if and only if $\mathscr{C}$ is an $R$-submodule of $R^n$.

\vskip 3mm \par
Quasi-cyclic (QC) codes are an important class of linear codes and have some good algebra structures [2-7, 16-18]. Recently, there are some research papers about QC codes over finite chain rings [2,3,5,7,10,17,22]. In [2], Aydin et al. studied QC codes over $\mathbb{Z}_4$ and obtained some new binary codes using the usual Gray map. Moreover, they characterized cyclic codes corresponding to free modules in terms of their generator polynomials. Bhaintwal et al. discussed QC codes over the prime integer residue ring $\mathbb{Z}_q$ [3]. They viewed a QC code of length $m\ell$ with index $\ell$ as an $\mathbb{Z}_q[x]/\langle x^m-1\rangle$-submodule of ${\rm GR}(q,\ell)[x]/\langle x^m-1\rangle$, where ${\rm GR}(q,\ell)$ was the $\ell$-th Galois extension ring of $\mathbb{Z}_q$. A sufficient condition for $1$-generator QC code to be $\mathbb{Z}_q$-free was given and some distance bounds for $1$-generator QC codes were also discussed. In [7], Cui et al. considered $1$-generator QC codes over $\mathbb{Z}_4$. Under some conditions, they gave the enumeration of quaternary $1$-generator QC codes of length $m\ell$ with index $\ell$, and described an algorithm to obtain one and only one generator for each $1$-generator QC code. Based on the idea in [7], we studied $1$-generator QC codes over another special finite chain ring $\mathbb{F}_q+u\mathbb{F}_q$ [10]. More recently, Cao used polynomial theory presented in [20] to study $1$-generator QC codes over arbitrary finite chain rings [5]. He discussed the parity check polynomials of $1$-generator QC codes and gave an explicit enumerator and algorithm for finding the generator of $1$-generator QC codes with some fixed parity check polynomial. However, it should be noted that all research papers discussed above base on one fact that the block length of the QC code is coprime with the characteristic of finite chain ring. In [22], Siap, et al. studied $1$-generator QC codes of arbitrary lengths over finite chain ring $\mathbb{F}_2+u\mathbb{F}_2$. They gave the generating set and the free condition of the $1$-generator QC code. Using Gary map, they also got some optimal binary linear codes over finite field $\mathbb{F}_2$.

\vskip 3mm \par
In this paper, we mainly consider QC codes over general finite chain rings. The paper is organized as follows. In section 2, we sketch some well known structural properties of cyclic codes over finite chain rings. Moreover, we discuss trace representations of cyclic codes. Some further results are obtained in this section (Proposition 2.7 and Theorem 2.8). In section 3, we discuss module structures of QC codes over finite chain rings, which are generalizations of QC codes over finite fields. This point of view for studying QC codes could give a lower bound on the minimum Hamming distance (Theorem 3.1) and a construct method of linear codes over finite fields (Theorem 3.2). In section 4, we discuss the trace representation of QC codes over finite chain rings, which lead to another lower bound on the minimum Hamming distance (Theorem 4.3). In section 5, we investigate $1$-generator QC codes. We give the structure of annihilator of the $1$-generator QC code (Theorem 5.1) and the free condition for $1$-generator QC code to be free (Theorem 5.2). In section 6, we give a sufficient and necessary condition for the $1$-generator QC code to be unique (Theorem 6.1). Under some conditions and using another method different from the point of view presented in [5], we also give an explicit enumeration formula for $1$-generator QC codes with a fixed annihilator (Theorem 6.4). In section 7, we describe how to obtain one and only one generator for each $1$-generator QC code (Theorem 7.2). Finally, we give an example to illustrate the main work in section 6 and section 7.

\vskip 3mm \noindent {\bf 2 Cyclic codes } \vskip 6mm \noindent
Let $T$ be a \emph{cyclic shift operator} $T$: $R^n\rightarrow R^n$, which transforms $v=(v_0, v_1, \ldots, v_{n-1})$ into $vT=(v_{n-1}, v_0, \ldots, v_{n-2})$. A linear code $\mathscr{C}$ is called the \emph{cyclic code} of length $n$ if it is invariant under $T$. In this paper, we assume $n$ to be a positive integer not divisible by the characteristic of the residue field $\mathbb{F}=\mathbb{F}_q$, so that $x^n-1$ is square free in $\mathbb{F}_q[x]$. Therefore $x^n-1$ has a unique decomposition as a product of monic basic irreducible pairwise coprime polynomials in $R[x]$. Let $f(x)$ be a factor of $x^n-1$ over $R$. Denote $\widehat{f}(x)=(x^n-1)/f(x)$. It is well known that the cyclic code of length $n$ over $R$ can be regarded as an ideal of $R[x]/\langle x^n-1\rangle$. In the following, we list the \emph{Chinese Reminder Theorem} (CRT) first, which will be used in other sections.

\vskip 3mm \noindent {\bf Theorem 2.1} (cf. [23] Theorem 2.9)\emph{ Let $f_1,f_2,\ldots,f_r$ be pairwise coprime monic polynomials of degree $\geqslant 1$ over $R$, $f=f_1f_2\ldots f_r$ and ${\mathcal R}_f=R[x]/\langle f\rangle$. Let $\widehat{f}_i=f/f_i$. Then there exist $a_i$, $b_i\in R[x]$ such that $a_if_i+b_i\widehat{f}_i=1$. Let $e_i=b_i\widehat{f}_i+\langle f\rangle$. Then
\vskip 1mm \noindent $(1)$ $e_1,e_2,\ldots,e_r$ are mutually orthogonal non-zero idempotents of ${\mathcal R}_f$;
\vskip 1mm \noindent $(2)$ $1=e_1+e_2+\cdots+e_r$ in  ${\mathcal R}_f$;
\vskip 1mm \noindent $(3)$ Let ${\mathcal R}_fe_i=\langle e_i\rangle$ be the principal ideal of ${\mathcal R}_f$ generated by $e_i$. Then $e_i$ is the identity of  ${\mathcal R}_fe_i$ and  ${\mathcal R}_fe_i=\langle\widehat{f}_i+\langle f\rangle\rangle$;
\vskip 1mm \noindent $(4)$ ${\mathcal R}_f=\bigoplus _{i=1}^r{\mathcal R}_fe_i$;
\vskip 1mm \noindent $(5)$ The map $R[x]/\langle f_i\rangle\rightarrow {\mathcal R}_fe_i$ defined by $g+\langle f_i\rangle\mapsto \langle g+\langle f\rangle\rangle e_i$ is a well-defined isomorphism of rings;
\vskip 1mm \noindent $(6)$ ${\mathcal R}_f=R[x]/\langle f\rangle\simeq \bigoplus_{i=1}^rR[x]/\langle f_i\rangle$.}            \hfill $\Box$

\vskip 3mm For the structure of the cyclic code over finite chain ring $R$, there have been some well known results as follows.

\vskip 3mm \noindent {\bf Theorem 2.2} (cf. [9] Theorem 3.2, 3.4, 3.6) \emph{Let $x^n-1=f_1f_2\cdots f_r$ be a representation of $x^n-1$ as a product of monic basic irreducible pairwise coprime polynomials over $R$. Then
\vskip 1mm \noindent $(1)$ Any ideal of $R[x]/\langle x^n-1\rangle$ is a sum of ideals of the form $\gamma^j\widehat{f}_i+\langle x^n-1\rangle$, where $0\leq j \leq s$, $1\leq i \leq r$.
\vskip 1mm \noindent $(2)$ Let $\mathscr{C}$ be a cyclic code over $R$. Then there exist a unique family of pairwise coprime monic polynomials $F_0$, $F_1$, $\ldots$ , $F_s$ in $R[x]$ (possibly, some of them are equal to $1$) such that $F_0F_1\cdots F_s=x^n-1$ and $\mathscr{C}=\langle\widehat{F}_1, \gamma\widehat{F}_2, \ldots, \gamma^{s-1}\widehat{F}_s\rangle$.
\vskip 1mm \noindent $(3)$ $R[x]/\langle x^n-1\rangle$ is a principal ideal ring. Let $\mathscr{C}$ be a cyclic code with notation in $(2)$. Then ${\mathscr{C}}=\langle\widehat{F}_1+\gamma\widehat{F}_2+\cdots+\gamma^{s-1}\widehat{F}_{s-1}\rangle$. }             \hfill $\Box$

\vskip 3mm \par  Let $\mathscr{C}$ be a cyclic code of length $n$ generated by $g(x)$ over $R$. Unlike the case over finite fields, $g(x)$ may be not a divisor of $x^n-1$. It is related to whether $\mathscr{C}$ is a free $R$-module or not.

\vskip 3mm \noindent {\bf Theorem 2.3 }(cf. [20] Proposition 4.11) \emph{Let $\mathscr{C}$ be a linear code over finite chain ring $R$. Then the following properties are equivalent
\vskip 1mm \noindent $(1)$ $\mathscr{C}$  is the Hensel lift of a cyclic code over $\overline{R}$;
\vskip 1mm \noindent $(2)$ $\mathscr{C}$ is a cyclic code and free;
\vskip 1mm \noindent $(3)$ There exists a polynomial $g(x)\in R[x]$ such that $\mathscr{C}=\langle g(x)\rangle$ and $g(x)|x^n-1$;
\vskip 1mm \noindent $(4)$ There exists a monic polynomial $g(x)\in R[x]$ such that $\{g(x)\}$ is the generating set in standard form for $\mathscr{C}$;
\vskip 1mm \noindent $(5)$ The dual code of $\mathscr{C}$ is the Hensel lift of a cyclic code over $\overline{R}$.} \hfill $\Box$

\vskip 3mm \par By Theorem 2.3, we can get the following proposition immediately.

\vskip 3mm \noindent {\bf Proposition 2.4 } \emph{A nonzero cyclic code $\mathscr{C}$ of length $n$ over $R$ is a free $R$-module if and only if it is generated by a polynomial $g(x)$ dividing $x^n-1$ over $R$. Moreover, if $\mathscr{C}$ is a free cyclic code, then the rank of $\mathscr{C}$ is $n-{\rm deg}g(x)$ and the set $\{g(x), xg(x),\ldots, x^{n-{\rm deg}g(x)-1}g(x)\}$ forms an $R$-basis for $\mathscr{C}$.}             \hfill $\Box$

\vskip 3mm \par  Cyclic code $\mathscr{C}$ of length $n$ over $R$ can also be described in terms of primitive roots of unity. Let $\xi^{i_1}$,$\xi^{i_2}$, $\ldots$, $\xi^{i_k}$ be $n$-th primitive roots of unity in some Galois extension ring ${\mathcal R}$ of $R$. Then the corresponding cyclic code $\mathscr{C}$ can be defined as $\mathscr{C}= \{c(x)\in R[x]/\langle x^n-1\rangle |~c(\xi^{i_j})=0, 1\leq j \leq k\}$. The generator polynomial $g(x)$ of $\mathscr{C}$ is the least common multiple of the minimal polynomials of $\xi^{i_j}$, $1\leq j \leq k$. Obviously, $g(x)$ is a monic divisor of $x^n-1$ over $R$. From Proposition 2.4, $\mathscr{C}$ is free.

 \vskip 3mm In the following, we consider the BCH-type bound of the cyclic code over finite chain ring $R$. The process of proof is analogous to that of the cyclic code over finite field. Here we omit it.

 \vskip 3mm \noindent {\bf Proposition 2.5 } \emph{Let $g(x)$ be the generator polynomial of the free cyclic code $\mathscr{C}$ of length $n$ and a monic divisor of $x^n-1$ over $R$. Suppose $g(x)$ has roots $\xi^b$, $\xi^{b+1}$, $\ldots$ , $\xi^{b+\delta-2}$, where $\xi$ is an $n$-th primitive root of unity in some Galois extension ring of $R$. Then the minimum Hamming distance of $\mathscr{C}$ is $d(\mathscr{C})\geq\delta$.}               \hfill $\Box$

 \vskip 3mm \noindent
{\bf Example 2.6} Let $R=\mathbb{F}_2+u\mathbb{F}_2$. It is well known that the polynomial $f(x)=x^3+x+1$ is an irreducible and primitive factor of $x^7-1$ over $\mathbb{F}_2$. Since $\mathbb{F}_2[x]$ is a subring of $R[x]$, we can regard $f(x)$ as a polynomial over $R$, i.e., the trivial Hensel lift of $f(x)\in \mathbb{F}_2[x]$ to $R[x]$. Denote ${\mathcal R}=R[x]/\langle f(x)\rangle$. Let $\xi$ be a root of $f(x)$. Then $\xi$ is a basic primitive element in ${\mathcal R}$, i.e., an element of order $2^3-1=7$ in ${\mathcal R}$. Suppose the generator polynomial of a cyclic code $\mathscr{C}$ of length $7$ over $R$ be defined by $g(x)={\rm lcm} (M_0(x), M_1(x), M_2(x))$, where $M_0(x)$, $M_1(x)$, $M_2(x)$ are minimal polynomials of $1$, $\xi$ and $\xi^2$, respectively. Then $M_0(x)=x+1$, $M_1(x)=M_2(x)=x^3+x+1$ implying $g(x)=(x+1)(x^3+x+1)=x^4+x^3+x^2+1$. Clearly, $g(x)$ is a monic divisor of $x^7-1$ over $R$. Therefore, from Proposition 2.4, $\mathscr{C}$ is free over $R$ and the set $\{g(x), xg(x),x^2g(x)\}$ forms an $R$-basis of $\mathscr{C}$. By Proposition 2.5, we have that the minimum Hamming distance of $\mathscr{C}$ is $d(\mathscr{C})\geq 4$. The generator matrix for $\mathscr{C}$ is given as follows
\begin{equation} \left(
  \begin{array}{ccccccc}
    1 & 0 & 1& 1 & 1 & 0 & 0\\
    0 & 1 & 0& 1 & 1 & 1 & 0\\
    0 &0 & 1& 0& 1 & 1& 1\\
  \end{array}
\right).
\end{equation}
Since $g(x)$ has the Hamming weight $4$, the minimum Hamming distance of $\mathscr{C}$, $d(\mathscr{C})=4$ actually. Thus $\mathscr{C}$ is a free cyclic code over $R$ with parameters $[7,3,4]$.

\vskip 3mm Let $x^n-1=f_1f_2\cdots f_r$, where each $f_i$ is a monic basic irreducible polynomial over finite chain ring $R$, $i=1,2,\ldots,r$. Suppose that $\xi$ is an $n$-th primitive root of unity and ${\mathcal R}$ is the smallest Galois extension ring of $R$, which contains the $n$-th primitive root of unity $\xi$. Therefore $x^n-1=(x-1)(x-\xi)\cdots (x-\xi^{n-1})$ over ${\mathcal R}$. Define the map $\pi$ as follows $$\pi :~{\mathcal R}[x]/\langle x^n-1\rangle\rightarrow \bigoplus_{i=0}^{n-1}{\mathcal R}[x]/\langle x-\xi^i\rangle$$  $$c(x)=c_0+c_1x+\cdots +c_{n-1}x^{n-1}\mapsto (c(1), c(\xi), \ldots, c(\xi^{n-1})).$$ If $c(x)\in R[x]/\langle x^n-1\rangle$, then from Theorem 2.1, we can deduce $\pi$ is an $R[x]$-module homomorphism. Denote $c(\xi^i)=A_i$ and $A(z)=\sum_{i=0}^{n-1}A_iz^{n-i}$. The $A(z)$ is called \emph{Mattsom-Solomon polynomial} associated with $c(x)$. Clearly,
\begin{equation} (A_0, A_1, \ldots, A_{n-1})=(c_0, c_1, \ldots, c_{n-1})\left(
  \begin{array}{cccc}
    1 & 1 & \ldots& 1 \\
    1 & \xi & \ldots& \xi^{n-1} \\
    \vdots &\vdots & \vdots& \vdots\\
    1 & \xi^{n-1} & \ldots& \xi^{(n-1)^2} \\
  \end{array}
\right).
\end{equation}
For this reason $A(z)$ is sometimes called a \emph{discrete Fourier transform} of $c(x)$. The inverse transform is given by $$c_j=\frac{1}{n}\sum_{k=0}^{n-1}A_k\xi^{-jk},~j=0,1,\ldots,n-1.$$

\vskip 3mm Suppose that ${\mathcal R}$ is an $m$-th Galois extension ring of finite chain ring $R$. It is well known that ${\mathcal R}$ is also with maximal ideal $\langle \gamma\rangle$ and the residue field $\mathbb{F}={\mathcal R}/\langle \gamma\rangle$ is $\mathbb{F}_{q^m}$. Every element $r$ of ${\mathcal R}$ can also be expressed uniquely in the form $$r=r_0+r_1\gamma+\cdots +r_{s-1}\gamma^{s-1},$$ where $r_0, r_1, \ldots, r_{s-1}$ belong to the Teichm\"{u}ller set $\mathcal {T}=\{0,1, \zeta, \ldots, \zeta^{q^m-2}\}$, where $\zeta$ is a $(q^m-1)$-th basic primitive element in ${\mathcal R}$. Define the Frobenius map $\phi$ on ${\mathcal R}$ to be the map induced by the map $r_0+r_1\gamma+\cdots +r_{s-1}\gamma^{s-1}\mapsto r_0^q+r_1^q\gamma+\cdots+r_{s-1}^q\gamma^{s-1}$, acting as the identity on $R$. Since the degree of the extension ${\mathcal R}$ over $R$ is $m$, $\phi^m$ is the identity map. The set of automorphism of ${\mathcal R}$ over $R$ forms a group with respect to the composition of maps, which is called the Galois group of ${\mathcal R}$ over $R$ and is denoted by ${\rm Gal}({\mathcal R}/R)$. It is well known that ${\rm Gal}({\mathcal R}/R)=\langle \phi\rangle$. For any $r\in {\mathcal R}$, we define the trace of $r$ to be $$Tr_{{\mathcal R}/R}(r)=r+\phi(r)+\cdots +\phi^{m-1}(r).$$ Since $\phi^i(r)=r_0^{q^i}+r_1^{q^i}\gamma+\cdots+r_{s-1}^{q^i}\gamma^{s-1}$, we have $$Tr_{{\mathcal R}/R}(r)=Tr_{{\mathcal R}/R}(r_0)+Tr_{{\mathcal R}/R}(r_1)\gamma+\cdots+Tr_{{\mathcal R}/R}(r_{s-1})\gamma^{s-1}.$$ By Hensel lift, there is a one-to-one correspondence between factors of $x^n-1$ and the $q$-cyclotomic cosets of $\mathbb{Z}_n$. Denote by $U_i~(1\leq i \leq r )$ the cyclotomic coset correspondence to $f_i$. Let ${\mathcal R}_i$ be the Galois extension ring of $R$ corresponding to the basic irreducible polynomial $f_i$, i.e., ${\mathcal R}_i=R[x]/\langle f_i\rangle$. Then for a fixed $u_i\in U_i$, we have $$nc_j=\sum_{i=1}^rTr_{{\mathcal R}_i/R}(A_i\xi^{-ju_i}).$$ Sometimes this is called the \emph{trace representation }of the cyclic code over finite chain ring $R$.

 \vskip 3mm In the following, we give a slight different trace representation of the cyclic code over finite chain ring $R$.

\vskip 3mm \noindent {\bf Proposition 2.7 } Let $\mathscr{C}$ be a free cyclic code of length $n$ over finite chain ring $R$. Suppose that non-negative integers $i_1, i_2, \ldots, i_k$ are in different $q$-cyclotomic cosets in $\mathbb{Z}_n$. Let $\xi$ be an $n$-th primitive root of unity and $\xi^{i_1}, \xi^{i_2}, \ldots, \xi^{i_k}$ be roots of the polynomial $m(x)=\prod_{j=1}^kM_j(x)$, where $m(x)$ is the generator polynomial of $\mathscr{C}^\perp$ and $M_j(x)$ is the minimal polynomial of $\xi^{i_j}$ over $R$. Then for any codeword $c(x)=c_0+c_1x+\cdots +c_{n-1}x^{n-1}$ of $\mathscr{C}$, we have $$c_v=\sum_{j=1}^kTr_{{\mathcal R}/R}a_j\xi^{vi_j},$$ where $a_j\in {\mathcal R}$, $v=0,1,\ldots, n-1$, and ${\mathcal R}$ is the smallest Galois extension ring of $R$ containing the $n$-th primitive root of unity $\xi$.

\vskip 3mm \noindent\emph{ Proof} Let $k=1$. Consider the following set $${\mathcal C}=\{(c_0, c_1, \ldots, c_{n-1})\in R^n|~c_v=Tr_{{\mathcal R}/R}a_j\xi^{vi_1}, v=0,1,\ldots, n-1\}.$$ Obviously, ${\mathcal C}$ is a nonzero linear code of length $n$ over $R$. If $c_{a_j}(x)=\sum_{v=0}^{n-1}Tr_{{\mathcal R}/R}a_j\xi^{vi_1}x^v$, then $c_{a_j\xi^{-i_1}}(x)=c_{a_j}(x)x$ in $R[x]/\langle x^n-1\rangle$ implying that ${\mathcal C}$ is cyclic. On the other hand the free cyclic code $\langle M_1(x)\rangle$ is contained in the dual code ${\mathcal C}^\perp$ of ${\mathcal C}$, which implies that $\langle M_1(x)\rangle^\perp \supseteq {\mathcal C}$. It should be noted that $\langle M_1(x)\rangle^\perp$ is a minimal free cyclic code with rank equality to the degree of $M_1(x)$, i.e., the minimal polynomial of $\xi^{i_1}$ over $R$. Since $R$ is a principal ideal ring, the cyclic code ${\mathcal C}$ is also free over $R$ implying that ${\mathcal C}=\langle M_1(x)\rangle^\perp$.
 \vskip 1mm For $k\geq 2$, using the facts that any free cyclic code is the direct sum of some minimal free cyclic codes, we can get the result in this proposition.                                    \hfill $\Box$

\vskip 3mm It is easy to see that $c_v=0$ if each $a_j=0$, $j=1,2,\ldots,k$. But sometimes $c_v$ may be identical zero even if there is $\{l_1, l_2, \ldots, l_d\}\subseteq \{1,2,\ldots,k\}$ such that $a_{l_z}\neq 0$, $z=1,2,\ldots,d$. Therefore we could ask a question that when $c_v$ is zero except the case $a_j=0$ for each $j=1,2,\ldots,k$? In the following we give an answer about this question.

\vskip 3mm \noindent {\bf Theorem 2.8 } \emph{Let $U_{vi_j}$ be a $q$-cyclomotic coset containing $vi_j$ mod $n$ for each $j=1,2,\ldots,k$. Let $a_1, a_2, \ldots, a_k\in {\mathcal R}\setminus \{0\}$. Then $c_v=0$ if and only if $|U_{vi_j}|=\tau_{vi_j}\neq m$ and $Tr_{{\mathcal R}/\widetilde{{\mathcal R}}_j}(a_j)=0$, where $\widetilde{{\mathcal R}}_j$ is the $\tau_{vi_j}$-th Galois extension ring of $R$ for all $j=1,2,\ldots,k$.}
\vskip 3mm \noindent\emph{ Proof} Firstly, we will prove $c_v=0$ if and only if $Tr_{{\mathcal R}/R}a_j\xi^{vi_j}=0$ for all $j=1,2,\ldots,k$. Let $a_j=a_{j0}+a_{j1}\gamma+\cdots+a_{j,s-1}\gamma^{s-1}$, where $a_{jg}\in {\mathcal T}=\{0,1,\zeta,\ldots,\zeta^{q^m-2}\}$, $\zeta$ is a basic primitive element in ${\mathcal R}$, $j=1,2,\ldots,k$ and $g=0,1,\ldots,s-1$. Then $c_v=0$ if and only if $\sum_{j=1}^kTr_{{\mathcal R}/R}a_j\xi^{vi_j}=\sum_{j=1}^kTr_{{\mathcal R}/R}(a_{j0}\xi^{vi_j})+\gamma\sum_{j=1}^kTr_{{\mathcal R}/R}(a_{j1}\xi^{vi_j})+\cdots+\gamma^{s-1}\sum_{j=1}^kTr_{{\mathcal R}/R}(a_{j,s-1}\xi^{vi_j})=0$ if and only if $\sum_{j=1}^kTr_{{\mathcal R}/R}(a_{jg}\xi^{vi_j})=0$ for all $g=0,1,\ldots,s-1$ if and only if $Tr_{{\mathcal R}/R}(a_{jg}\xi^{vi_j})=0$ for all $j=1,2,\ldots,k$ and $g=0,1,\ldots,s-1$ if and only if $Tr_{{\mathcal R}/R}a_j\xi^{vi_j}=0$ for all $j=1,2,\ldots,k$.

\vskip 1mm Secondly, we will prove $Tr_{{\mathcal R}/R}a_j\xi^{vi_j}=0$ if and only if $|U_{vi_j}|=\tau_{vi_j}\neq m$ and $Tr_{{\mathcal R}/\widetilde{{\mathcal R}}_j}(a_j)=0$. Since $\tau_{vi_j}$ necessarily divides $m$, $\widetilde{{\mathcal R}}$ is a subring of ${\mathcal R}$. Therefore $Tr_{{\mathcal R}/\widetilde{{\mathcal R}}}$ makes sense. From $a_j\in {\mathcal R}\setminus \{0\}$, we have $|U_{vi_j}|\neq m$. By Theorem 2.2 in [12], we deduce that there are $q^{(m-\tau_{vi_j})s}$ $a_j$'s in ${\mathcal R}$ such that $Tr_{{\mathcal R}/R}a_j\xi^{vi_j}=0$. The number of elements in the kernel of $Tr_{{\mathcal R}/\widetilde{{\mathcal R}}}$ is also $q^{ms}/q^{\tau_{vi_j}s}=q^{(m-\tau_{vi_j})s}$. For any $b_j$ in this kernel, we have $$Tr_{{\mathcal R}/R}(b_j\xi^{vi_j})=Tr_{\widetilde{{\mathcal R}}/R}(Tr_{{\mathcal R}/\widetilde{{\mathcal R}}}(b_j\xi^{vi_j}))=Tr_{\widetilde{{\mathcal R}}/R}(\xi^{vi_j}Tr_{{\mathcal R}/\widetilde{{\mathcal R}}}(b_j))=0.$$ Thus we have $a_j$ must be in the kernel of $Tr_{{\mathcal R}/\widetilde{{\mathcal R}}}$. Conversely, reading the above equality from left to right, replacing $b_j$ by $a_j$, proves the claim.                \hfill $\Box$

\vskip 3mm \noindent {\bf 3 Module structure of quasi-cyclic codes} \vskip 6mm \noindent

Let $T$ be a cyclic shift operator $T$: $R^N\rightarrow R^N$, which transforms $v=(v_0, v_1, \ldots, v_{N-1})$ into $vT=(v_{N-1}, v_0, \ldots, v_{N-2})$. A linear code $\mathscr{C}$ is called \emph{quasi-cyclic} (QC) code if it is invariant under $T^{\ell}$ for some positive integer $\ell$. The smallest $\ell$ such that $T^\ell(\mathscr{C})=\mathscr{C}$ is called the index of $\mathscr{C}$. Clearly, $\ell$ is a divisor of $N$. Let $N=n\ell$. Define an $R$-module isomorphism as follows $$\rho:~R^{n\ell}\rightarrow (R[x]/\langle x^n-1\rangle)^\ell$$ $$(v_{00}, v_{01},\ldots,v_{0,\ell-1}, v_{10}, v_{11}, \ldots,v_{1,\ell-1},\ldots,v_{n-1,0}, v_{n-1,1}, \ldots, v_{n-1,\ell-1} )$$ $$\mapsto (v_0(x), v_1(x), \ldots, v_{\ell-1}(x)),$$ where $v_i(x)=\sum_{j=0}^{n-1}v_{ji}x^j$, $i=0,1,\ldots,\ell-1$. Then the QC code $\mathscr{C}$ is equivalent to saying that, for any $(v_0(x), v_1(x), \ldots, v_{\ell-1}(x))\in \rho(\mathscr{C})$, $(xv_0(x), xv_1(x), \ldots, xv_{\ell-1}(x))\in\rho(\mathscr{C})$. Therefore, $\mathscr{C}$ is a QC code of length $n\ell$ with index $\ell$ if and only if $\rho(\mathscr{C})$ is an $R[x]/\langle x^n-1\rangle$-submodule of $(R[x]/\langle x^n-1\rangle)^\ell$. This definition of the QC code is known as conventional row circulant. But in this section, we generalize another definition of the QC code in [16] to finite chain ring $R$.

\vskip 3mm Let $v=(v_{00}, v_{01},\ldots,v_{0,\ell-1}, v_{10}, v_{11}, \ldots,v_{1,\ell-1},\ldots,v_{n-1,0}, v_{n-1,1}, \ldots, v_{n-1,\ell-1} )\in R^{n\ell}$. Define an isomorphism between $R^{n\ell}$ and ${\mathcal R}^n$ by associating with each $\ell$-tuple $(v_{i0}, v_{i1}, \ldots,v_{i,\ell-1})$, $i=0,1,\ldots,n-1$, and the element $v_i\in {\mathcal R}$ represented as $v_i=v_{i0}+v_{i1}\xi+\cdots+v_{\ell-1}\xi^{\ell-1}$, where the set $\{1, \xi, \xi^2, \ldots, \xi^{\ell-1}\}$ forms an $R$-basis of ${\mathcal R}$. Then every element in $R^{n\ell}$ is one-to-one correspondence with an element in ${\mathcal R}^n$. The operator $T^\ell$ for some element $(v_{00}, v_{01},\ldots,v_{0,\ell-1},\ldots, \\v_{n-1,0}, v_{n-1,1}, \ldots,v_{n-1,\ell-1} )\in R^{n\ell}$ corresponds to the element $(v_{n-1}, v_0,\ldots, v_{n-2})\in {\mathcal R}^n$. Indicating the block positions with increasing powers of $x$, the vector $v\in R^{n\ell}$ can be associated with the polynomial $v_0+v_1x+\cdots+v_{n-1}x^{n-1}\in {\mathcal R}[x]$. An $R[x]/\langle x^n-1\rangle$-module isomorphism between $R^{n\ell}$ and ${\mathcal R}[x]/\langle x^n-1\rangle$, which is defined as $\psi(v)=v_0+v_1x+\cdots+v_{n-1}x^{n-1}$. In this setting, multiplication by $x$ of any element of ${\mathcal R}[x]/\langle x^n-1\rangle$ is equal to applying $T^\ell$ to operate the element of $R^{n\ell}$. It follows that there is a one-to-one correspondence between $R[x]/\langle x^n-1\rangle$-submodule of ${\mathcal R}[x]/\langle x^n-1\rangle$ and the QC code of length $n\ell$ with index $\ell$ over $R$. In addition, let $\mathscr{C}$ be a QC code of length $n\ell$ with index $\ell$ over $R$. It is also can be regarded as an $R$-submodule of ${\mathcal R}[x]/\langle x^n-1\rangle$ because of the equivalence of $R^{n\ell}$ and ${\mathcal R}[x]/\langle x^n-1\rangle$.

\vskip 3mm \par Let $\mathscr{C}$ be a QC code of length $n\ell$ with index $\ell$ over $R$, and generated by elements $v_1(x), v_2(x),\ldots, v_r(x)\in {\mathcal R}[x]/\langle x^n-1\rangle$ as an $R[x]/\langle x^n-1\rangle$-submodule of ${\mathcal R}[x]/\langle x^n-1\rangle$. Then $\mathscr{C}=\{a_1(x)v_1(x)+a_2(x)v_2(x)+\cdots+a_r(x)v_r(x)|~a_i(x)\in R[x]/\langle x^n-1 \rangle, i=1,2,\ldots,r\}$. As discussed above, $\mathscr{C}$ is also an $R$-submodule of ${\mathcal R}[x]/\langle x^n-1\rangle$. For an $R$-submodule of ${\mathcal R}[x]/\langle x^n-1\rangle$, $\mathscr{C}$ is generated by the following set $\{v_1(x), xv_1(x), \ldots, x^{n-1}v_1(x),  \ldots, v_r(x), xv_r(x), \ldots, x^{n-1}v_r(x)\}$.

\vskip 3mm \par If $\mathscr{C}$ is generated by a single element $v(x)$ as an $R[x]/\langle x^n-1\rangle$-submodule of ${\mathcal R}[x]/\langle x^n-1\rangle$, then $\mathscr{C}$ is called the \emph{$1$-generator} QC code. Let the preimage of $v(x)$ in $R^{n\ell}$ be $v$. Then for the $1$-generator QC code $\mathscr{C}$, we have $\mathscr{C}$ is generated by the set $\{v, T^\ell v, \ldots, T^{\ell(n-1)}v\}$. It is the conventional of row circulant definition of $1$-generator QC code. In fact, let $v(x)=v_0+v_1x+\cdots+v_{n-1}x^{n-1}$ be a polynomial in ${\mathcal R}[x]/\langle x^n-1\rangle$, where $v_i=v_{i0}+v_{i1}\xi+\cdots+v_{i,\ell-1}\xi^{\ell-1}$, $i=0,1,\ldots,n-1$. Then $v(x)$
becomes an $\ell$-tuple of polynomials over $R$ each of degree at most $n-1$ with the fixed $R$-basis $\{1, \xi, \xi^2, \ldots,\xi^{\ell-1}\}$. Therefore, $v(x)$ becomes an element of $(R[x]/\langle x^n-1\rangle)^\ell$. So $\mathscr{C}$ is an $R[x]/\langle x^n-1\rangle$-submodule of $(R[x]/\langle x^n-1\rangle)^\ell$, i.e. the conventional row circulant definition of QC code.

\vskip 3mm \par Since $R[x]/\langle x^n-1\rangle$ is a subring of ${\mathcal R}[x]/\langle x^n-1\rangle$ and $\mathscr{C}$ is an $R[x]/\langle x^n-1\rangle$-submodule of ${\mathcal R}[x]/\langle x^n-1\rangle$, it is in particular a submodule of an ${\mathcal R}[x]/\langle x^n-1\rangle$-submodule of ${\mathcal R}[x]/\langle x^n-1\rangle$, i.e. the cyclic code $ \widetilde{\mathscr{C}}$ of length $n$ over ${\mathcal R}$. Therefore, $$d(\mathscr{C})\geq d(\widetilde{\mathscr{C}}),$$ where $d(\mathscr{C})$ and $d(\widetilde{\mathscr{C}})$ are minimum Hamming distances of $\mathscr{C}$ and $\widetilde{\mathscr{C}}$, respectively.

\vskip 3mm \noindent {\bf Theorem 3.1} \emph{Let $\mathscr{C}$ be an $r$-generator QC code of length $n\ell$ with index $\ell $ over $R$ and generated by the set $\{v_1(x), v_2(x), \ldots,v_r(x)\}$, where $v_i(x)\in {\mathcal R}[x]/\langle x^n-1\rangle$, $i=1,2,\ldots,v$. Then $\mathscr{C}$ has a lower bound on the minimum Hamming distance given by $$d(\mathscr{C})\geq d(\widetilde{\mathscr{C}})d(\mathscr{B}),$$ where $\widetilde{\mathscr{C}}$ is a cyclic code of length $n$ over ${\mathcal R}$ with generator polynomials $v_1(x), v_2(x), \ldots, v_r(x)$, and $\mathscr{B}$ is a linear code of length $\ell$ generated by the set $\{{\mathcal V}_{ij}, i=1,2,\ldots,r, j=0,1,\ldots,n-1\}\subseteq R^\ell$ where each ${\mathcal V}_{ij}$ is the vector equivalent of the $j$-th coefficient of $v_i(x)$ with respect to an $R$-basis $\{ 1, \xi, \ldots, \xi^{\ell-1}\}$.}

\vskip 3mm \noindent\emph{ Proof} The result follows from the Theorem 5 in [16].            \hfill $\Box$

\vskip 3mm  In the reset of this section, we give an application of the above discussion. This application leads to construct QC codes over finite fields.

\vskip 3mm Let $R=\mathbb{F}_q+u\mathbb{F}_q+\cdots+u^{\ell-1}\mathbb{F}_q$, where $u^\ell=0$ and $\ell$ is a positive integer. Consider a cyclic code $\widetilde{\mathscr{C}}$ of length $n$ generated by a polynomial $v(x)$ over $R$. Let $\mathscr{C}$ be a linear code of length $n\ell$ spanned by the set $\{v(x), xv(x), \ldots, x^{n-1}v(x)\}$ over $\mathbb{F}_q$. Then $\mathscr{C}$ is a $1$-generator QC code of length $n\ell$ with index $\ell$. If $v(x)=v_0+v_1x+\cdots+v_{n-1}x^{n-1} \in R[x]/\langle x^n-1\rangle$, then each $v_i$ is an $\ell$-tuple with respect to the fixed $\mathbb{F}_q$-basis $\{1,u,\ldots,u^{\ell-1}\}$ of $R$. Now let the set $\{v_0, v_1, \ldots, v_{n-1}\}$ generate a linear code $\mathscr{B}$ of length $\ell$ over $\mathbb{F}_q$. By Theorem 3.1, we have

\vskip 3mm \noindent{\bf Theorem 3.2} \emph{Let $\mathscr{C}$ be a quasi-cyclic code of length $n\ell$ with index $\ell$ over finite field $\mathbb{F}_q$ generated by the set $\{ v(x),xv(x),\ldots,x^{n-1}v(x) \}$, where  $v(x)=v_0+v_1x+\cdots+v_{n-1}x^{n-1}\in R[x]/\langle x^n-1\rangle$. Then
\vskip 1mm \noindent $(1)$ $\mathscr{C}$ has a lower bound on the minimum Hamming distance given by $d(\mathscr{C})\geq d(\widetilde{\mathscr{C}})d(\mathscr{B})$, where $\widetilde{\mathscr{C}}$ is a cyclic code of length $n$ over $R$ generated by the polynomial $g(x) \in R[x]/\langle x^n-1\rangle$, and $\mathscr{B}$ is a linear code of length $\ell$ generated by $\{v_0, v_1, \ldots, v_{n-1}\}$ where each $v_i$ is an $\ell$-tuple with respect to a fixed $\mathbb{F}_q$-basis $\{1,u,\ldots, u^{\ell-1}\}$ of $R$.}
\vskip 1mm \noindent $(2)$ \emph{If the cyclic code $\widetilde{\mathscr{C}}$ in $(1)$ is free and the generator polynomial $g(x)$ has $\delta-1$ consecutive roots in some Galois extension ring of $R$, and if the set $\{v_0, v_1, \ldots, v_{n-1}\}$ generates a cyclic code $\mathscr{B}$ over finite field $\mathbb{F}_q$ of length $\ell$ such that the generator polynomial of $\mathscr{B}$ has $\varepsilon-1$ consecutive roots in some Galois extension field of $\mathbb{F}_q$, then $d(\mathscr{C})\geq \delta\varepsilon$.}            \hfill $\Box$

\vskip 3mm \noindent {\bf Example 3.3} Let $R=\mathbb{F}_2+u\mathbb{F}_2+u^2\mathbb{F}_2$. Suppose $R=\{0,1,u,v,uv,u^2,v^2,v^3\}$, where $u^3=0$, $v=1+u$, $v^2=1+u^2$, $v^3=1+u+u^2$, $uv=u+u^2$. It is well known that $x^7-1=(x+v^3)(x^3+uvx^2+v^2x+v^3)(x^3+vx^2+ux+v^3)$, where $x+v^3$, $x^3+uvx^2+v^2x+v^3$ and $x^3+vx^2+ux+v^3$ are basic irreducible polynomials over $R$. Let ${\mathcal R}=R[x]/(x^3+uvx^2+v^2x+v^3)$. Since $x^3+uvx^2+v^2x+v^3$ is a basic primitive polynomial over $R$, the root $\xi$ of $x^3+uvx^2+v^2x+v^3$ is a primitive element in ${\mathcal R}$. Taking $v(x)=(x+v^3)(x^3+uvx^2+v^2x+v^3)=x^4+x^3+(1+u+u^2)x^2+u^2x+(1+u^2)$, then the cyclic code $\widetilde{\mathscr{C}}$ of length $7$ generated by $v(x)$ is free, and by Proposition 2.5, we have the minimum Hamming distance of $\widetilde{\mathscr{C}}$ at least $4$. The non-zero coefficients of $v(x)$ correspond to the elements $(1,0,1)$, $(0,0,1)$, $(1,1,1)$, $(1,0,0)$, $(1,0,0)$ with respect to the $\mathbb{F}_2$-basis $\{1,u,u^2\}$ of $R$ and they generate a cyclic code $\mathscr{B}$ of length $3$ with the minimum Hamming distance $1$ over $\mathbb{F}_2$. Therefore, $\mathscr{C}$ is a $1$-generator QC code of length $21$ with dimension $3$ and minimum Hamming distance at least $4\times1=4$ over finite field $\mathbb{F}_2$. A generator matrix for $\mathscr{C}$ is given as follows
\begin{equation} \left(
  \begin{array}{ccccccc}
    101 & 001 & 111& 100 & 100 & 000 & 000\\
    000 & 101 & 001& 111 & 100 & 100 & 000\\
    000 & 000 & 101& 001& 111 & 100& 100\\
  \end{array}
\right).
\end{equation}
In fact the minimum Hamming distance of $\mathscr{C}$ is $8$. Therefore, $\mathscr{C}$ is a QC code with parameters $[21, 3, 8]$ over $\mathbb{F}_2$.

\vskip 3mm \noindent {\bf 4 Trace representation of quasi-cyclic codes} \vskip 6mm \noindent

 Let $x^n-1=f_1f_2\ldots f_r$, where each $f_i$, $i=1,2,\ldots,r$, is the monic basic irreducible polynomial with degree $\ell_i$ over $R$. Then from Theorem 2.1, we have $(R[x]/\langle x^n-1\rangle)^\ell \simeq \bigoplus_{i=1}^\ell(R[x]/\langle f_1\rangle)^\ell$. Therefore if $\mathscr{C}$ is a QC code of length $n\ell$ with index $\ell$ over $R$, then $$\mathscr{C}=\bigoplus_{i=1}^r \mathscr{C}_i,$$ where $\mathscr{C}_i$, $i=1,2,\ldots,r$, is a linear code of length $\ell$ over the $\ell_i$-th Galois extension ring ${\mathcal R}_i$ of $R$. This is called the canonical decomposition of the QC code $\mathscr{C}$. The trace representation of cyclic code over finite chain $R$ can give the following characterization result on QC code as follows.

\vskip 3mm \noindent {\bf Theorem 4.1 } (cf. Theorem 5.1 [17]) \emph{Let $x^n-1=f_1f_2\ldots f_r$, where each $f_i$, $i=1,2,\ldots,r$, is the basic irreducible polynomial with degree $\ell_i$ over $R$. Denote ${\mathcal R}_i=R[x]/\langle f_i\rangle$. Let $U_i$ denote the $q$-cyclotomic coset mod $n$ corresponding to $f_i$. Fix a representatives $u_i\in U_i$ from each cyclotomic coset. Let $\mathscr{C}_i$ be a linear code of length $\ell$ over ${\mathcal R}_i$ for all $i=1,2,\ldots,r$. For $ \widetilde{\textbf{c}}_i\in \mathscr{C}_i$ and each $j=0,1,\ldots,n-1$, let the vector $\textbf{c}_j=\sum_{i=1}^rTr_{{\mathcal R}_i/R}(\widetilde{\textbf{c}}_i \xi^{-ju_i})$. Then the code $$\mathscr{C}=\{(\textbf{c}_0, \textbf{c}_1, \ldots, \textbf{c}_{n-1})|~\widetilde{\textbf{c}}_i\in \mathscr{C}_i\}$$ is a QC code of length $n\ell$ with index $\ell$ over $R$. Conversely, every QC code of length $n\ell$ with index $\ell$ over $R$ is obtained through this construction. }  \hfill $\Box$

\vskip 3mm  Let $R\subset \widetilde{R} \subset {\mathcal R}$ be Galois extension. If $\omega \in {\mathcal R}$ such that $Tr_{{\mathcal R}/\widetilde{R}}(\omega)=1$, then for any $\vartheta \in \widetilde{R}$ we have $$Tr_{\widetilde{R}/R}(\vartheta)=Tr_{{\mathcal R}/R}(\omega\vartheta).$$

\vskip 3mm \noindent {\bf Theorem 4.2 } \emph{Let $\mathscr{C}$ be a QC code defined as above. Let $\omega_1, \omega_2, \ldots, \omega_r \in {\mathcal R}$ be elements with $Tr_{{\mathcal R}/{\mathcal R}_i}(\omega_i)=1$ for all $i=1,2,\ldots,r$. Then
\vskip 1mm \noindent $(1)$~Any codeword $(\textbf{c}_0, \textbf{c}_1, \ldots, \textbf{c}_{n-1})\in \mathscr{C}$ is of the form $\textbf{c}_j=\sum_{i=1}^rTr_{{\mathcal R}/R}(\widetilde{\textbf{c}}_i \omega_i\xi^{-ju_i})$, for all $j=0,1,\ldots,n-1$.
\vskip 1mm \noindent $(2)$~The columns of any codeword $c\in \mathscr{C}$ lie in a free cyclic code $\mathscr{B}$ of length $n$ over $R$, which dual code $\mathscr{B}^\perp$ has roots $\xi^{-u_1}, \xi^{-u_2}, \ldots, \xi^{-u_r}$, where $\xi$ is an $n$-th primitive root of unity in ${\mathcal R}$;
\vskip 1mm \noindent $(3)$~ For any column $$\widehat{\textbf{c}}_\nu=(\Sigma_{i=1}^rTr_{{\mathcal R}/R}(\widetilde{c}_{i,\nu}\omega_i), \Sigma_{i=1}^rTr_{{\mathcal R}/R}(\widetilde{c}_{i,\nu}\omega_i\xi^{-u_i}), \ldots, \Sigma_{i=1}^rTr_{{\mathcal R}/R}(\widetilde{c}_{i,\nu}\omega_i\xi^{-(n-1)u_i}) ),$$ where $\widetilde{\textbf{c}}_i=(\widetilde{c}_{i1}, \widetilde{c}_{i2}, \ldots, \widetilde{c}_{i\ell}) \in {\mathcal R}_i$ and $\nu=1,2,\ldots,\ell$, we have $\widehat{\textbf{c}}_\nu=\textbf{0}$ if and only if $\widetilde{c}_{1,\nu}=\widetilde{c}_{2,\nu}=\cdots =\widetilde{c}_{r,\nu}=0$.}
\vskip 3mm \noindent\emph{ Proof} $(1)$~Clearly, $\sum_{i=1}^rTr_{{\mathcal R}/R}(\widetilde{\textbf{c}}_i \omega_i\xi^{-ju_i})=\sum_{i=1}^rTr_{{\mathcal R}_i/R}(\widetilde{\textbf{c}}_i\xi^{-ju_i})=\textbf{c}_j$;
\vskip 1mm \noindent $(2)$~For any column $$\widehat{\textbf{c}}_\nu=(\Sigma_{i=1}^rTr_{{\mathcal R}/R}(\widetilde{c}_{i,\nu}\omega_i), \Sigma_{i=1}^rTr_{{\mathcal R}/R}(\widetilde{c}_{i,\nu}\omega_i\xi^{-u_i}), \ldots, \Sigma_{i=1}^rTr_{{\mathcal R}/R}(\widetilde{c}_{i,\nu}\omega_i\xi^{-(n-1)u_i}) ),$$ the $v$-th component $$\widehat{c}_v=\sum_{i=1}^rTr_{{\mathcal R}/R}(\widetilde{c}_{i, \nu}\omega_i\xi^{-vu_i}),$$ where $\widetilde{\textbf{c}}_i=(\widetilde{c}_{i1}, \widetilde{c}_{i2}, \ldots, \widetilde{c}_{i\ell}) \in {\mathcal R}_i$ and $\nu=1,2,\ldots,\ell$. Since $\widetilde{c}_{i, \nu}\omega_i \in {\mathcal R}$, from Proposition 2.7, we have $\widehat{\textbf{c}}_\nu$ lies in a free cyclic code $\mathscr{B}$ of length $n$ over $R$, which dual code $\mathscr{B}^\perp$ has roots $\xi^{-u_1}, \xi^{-u_2}, \ldots, \xi^{-u_r}$;
\vskip 1mm \noindent $(3)$~$\widehat{\textbf{c}}_\nu=\textbf{0}$ if and only if each $v$-th component is zero for all $v=0,1,\ldots,n-1$ if and only if $\Sigma_{i=1}^rTr_{{\mathcal R}/R}(\widetilde{c}_{i,\nu}\omega_i\xi^{-vu_i})=0$ if and only if $Tr_{{\mathcal R}/R}(\widetilde{c}_{i,\nu}\omega_i\xi^{-vu_i})=0$ for all $i=1,2,\ldots,r$. \\
$(i)$~If $\ell_1=\ell_2=\cdots =\ell_r=m$, then $Tr_{{\mathcal R}/R}(\widetilde{c}_{i,\nu}\omega_i\xi^{-vu_i})=0$ for all $i=1,2,\ldots,r$ if and only if $\widetilde{c}_{1,\nu}=\widetilde{c}_{2,\nu}=\cdots =\widetilde{c}_{r,\nu}=0$;\\
$(ii)$~If there exists a set $\{j_1, j_2, \ldots, j_d\}\subseteq \{1,2,\ldots,r\}$ such that $\ell_{j_k}< m$ for all $k\in \{j_1, j_2, \ldots, j_d\}$ and $\ell_p=m$ for all $p\in \{1,2,\ldots,r\}\backslash \{j_1, j_2, \ldots, j_d\}$, then $Tr_{{\mathcal R}/R}(\widetilde{c}_{i,\nu}\omega_i\xi^{-vu_i})=0$ for all $i=1,2,\ldots,r$ if and only if $\widetilde{c}_{\ell_p, \nu}=0$ and $Tr_{{\mathcal R}/{\mathcal R}_i}(\widetilde{c}_{\ell_{j_k}}\omega_i)=\widetilde{c}_{\ell_{j_k}} Tr_{{\mathcal R}/{\mathcal R}_i}(\omega_i)=\widetilde{c}_{\ell_{j_k}}=0$. Therefore, we have proved $\widehat{\textbf{c}}_\nu=\textbf{0}$ if and only if $\widetilde{c}_{1,\nu}=\widetilde{c}_{2,\nu}=\cdots =\widetilde{c}_{r,\nu}=0$.          \hfill $\Box$

\vskip 3mm  As a consequence of Theorem 4.2, we can exhibit a minimum Hamming distance bound for the QC code. We assume that $$d(\mathscr{C}_1)\geq d(\mathscr{C}_2)\geq \cdots \geq d(\mathscr{C}_r).$$ For any nonempty sunset $\mathcal {I}=\{i_1, i_2, \ldots, i_t\} \subseteq \{1,2,\ldots,r\}$ with $1\leq i_1<i_2<\cdots <i_t\leq r$, let $\mathscr{B}_{\mathcal {I}}=\mathscr{B}_{i_1, i_2, \ldots, i_t}$ be a free cyclic code of length $n$ over $R$, which dual code $\mathscr{B}_{\mathcal {I}}^\perp$ has roots $\xi^{-u_{i_1}}, \xi^{-u_{i_2}}, \ldots, \xi^{-u_{i_t}}$. If $\emptyset \neq \mathcal {I}_1\subset \mathcal {I}_2\subseteq \{1,2,\ldots,r\}$, then $\mathscr{B}_{\mathcal {I}_1}\subset \mathscr{B}_{\mathcal {I}_2}$ and hence $d(\mathscr{B}_{\mathcal {I}_1})\geq d(\mathscr{B}_{\mathcal {I}_2})$.
\vskip 3mm For $\mathcal {I}$ defined above, we define
\begin{equation}
d_{\mathcal {I}}=d_{i_1, i_2, \ldots, i_t}= \begin{cases}
         d(\mathscr{C}_{i_1})d(\mathscr{B}_{i_1}) ~~~~~~~~~~~~~~~~~~~~~~~~~~~~~~~~~~~~~~~~~~~~~{\rm if}~ t=1 \\
                  (d(\mathscr{C}_{i_1})-d(\mathscr{C}_{i_2}))d(\mathscr{B}_{i_1}) +(d(\mathscr{C}_{i_2})-d(\mathscr{C}_{i_3}))d(\mathscr{B}_{i_1,i_2})\\
         ~~~~~~~~~~~~~~~~~~~~~~~~~~~~~~ \vdots  ~~~~~~~~~~~~~~~~~~~~~~~~~~~~~~~{\rm if}~t\geq2\\
          +(d(\mathscr{C}_{i_{t-1}})-d(\mathscr{C}_{i_t}))d(\mathscr{B}_{i_1,i_2, \ldots,i_{t-1}})+ d(\mathscr{C}_{i_t})d(\mathscr{B}_{i_1,i_2, \ldots,i_t})
                 \end{cases}.
\end{equation}

Let $\mathcal {J}=\mathcal {I}\backslash \{i_\mu\}$ for some $\mu \in \{2,3,\ldots,t\}$. Then $d_{\mathcal {J}}\geq d_{\mathcal {I}}$. (See Lemma 4.7 in [13].)

\vskip 3mm In the following, we give the minimum Hamming distance bound for QC codes over finite chain ring $R$, which is an interesting generalization of Theorem 4.8 in [13].

\vskip 3mm \noindent {\bf Theorem 4.3 }\emph{ Let $\mathscr{C}$ be a QC code as discussed above. Then the minimum Hamming distance of $\mathscr{C}$ satisfies $$d(\mathscr{C})\geq {\rm min}\{d_r, d_{r-1,r}, \ldots, d_{1,2,\ldots,r}\}.$$}

\vskip 3mm \noindent\emph{ Proof} Let $c$ be a nonzero codeword of $\mathscr{C}$. Suppose that $\widetilde{\textbf{c}}_{i_k}\in \mathscr{C}_i$ for all $k=1,2,\ldots,t$, where $\{i_1, i_2, \ldots, i_t\}\subseteq {1,2,\ldots,r}$ and $1\leq i_1< i_2<\cdots <i_t \leq r$. If $t=1$, then by Theorem 4.2 (2) there exists at least $d(\mathscr{C}_1)$ nonzero columns in $\mathscr{C}$ implying the minimum possible weight for such code is $d(\mathscr{C})\geq d(\mathscr{C}_{i_1})d(\mathscr{B}_{i_1})$. If $t\geq 2$, then the weight for such code $\mathscr{C}$ is minimized if $${\rm Supp}(\widetilde{\textbf{c}}_{i_t})\subseteq {\rm Supp}(\widetilde{\textbf{c}}_{i_{t-1}})\subseteq \cdots\subseteq {\rm Supp}(\widetilde{\textbf{c}}_{i_1}),$$ where ${\rm Supp}(\widetilde{\textbf{c}}_{i_k})$ denotes the nonzero coordinates of $\widetilde{\textbf{c}}_{i_k}$ for all $k=1,2,\ldots,t$. By the proof process of Theorem 4.5 in [13], the lowest possible weight for such code in this case is $$d(\mathscr{C})\geq d_{\mathcal {I}}=(d(\mathscr{C}_{i_1})-d(\mathscr{C}_{i_2}))d(\mathscr{B}_{i_1})+\cdots+d(\mathscr{C}_{i_t})d{\mathscr{B}_{i_1, i_2, \ldots, i_t}},$$ which implies that $$d(\mathscr{C})\geq {\rm min}\{ d_{\mathcal {I}}|~ \mathcal {I}=\{i_1, i_2, \ldots, i_t\}\subseteq \{1,2,\ldots,r\}~with~i_1< i_2<\cdots<i_t\}.$$  Let $\mathcal {N}\subseteq \{1,2,\ldots,r\}$ and let $i$ be the minimal element in $\mathcal {N}$. Adjoining one element at a time, we have $\mathcal {N}\subseteq \mathcal {N}_1\subseteq \cdots \subseteq \{i,i+1,\ldots,r\}$. Then $$d_{\mathcal {N}}\geq d_{\mathcal {N}_1}\geq \cdots\geq d_{i, i+1, \ldots, r}.$$ Hence, the minimum Hamming distance of $\mathscr{C}$  is equal to $$d(\mathscr{C})\geq {\rm min}\{d_r, d_{r-1,r}, \ldots, d_{1,2,\ldots,r}\}.$$                 \hfill $\Box$

\vskip 3mm \noindent {\bf Example 4.4 } Consider a QC code $\mathscr{C}$ of length $14$ with index $2$ generated by $(a_0(x), a_1(x))$ over $R=\mathbb{F}_2+u\mathbb{F}_2$, where $a_0(x)=x^4+x^2+x$ and $a_1(x)=x^4+x^3+x^2+x+1$. Since $(R[x]/\langle x^7-1\rangle)^2\cong (R[x]/\langle x-1\rangle)^2\bigoplus(R[x]/\langle x^3+x^2+1\rangle)^2\bigoplus(R[x]/\langle x^3+x+1\rangle)^2 $, we have $$\mathscr{C}=\bigoplus_{i=1}^3\mathscr{C}_i,$$ where $\mathscr{C}_1$ is a linear code of length $2$ generated by $(1,0)$ over $R$, $\mathscr{C}_2$ is a linear code of length $2$ generated by $(1, x^2+1)$ over $R[x]/\langle x^3+x^2+1\rangle$ and $\mathscr{C}_3$ is a zero code over $R[x]/\langle x^3+x+1\rangle$. Clearly, $d(\mathscr{C}_1)=d(\mathscr{C}_2)=1$, $d(\mathscr{C}_3)=0$. Hence Theorem 4.3 yields $$d(\mathscr{C})\geq {\rm min}\{6,3\}=3.$$

\vskip 3mm \noindent {\bf 5 $1$-generator quasi-cyclic codes} \vskip 6mm \noindent
Let $R$ be finite chain ring with maximal ideal $\langle \gamma\rangle$ and the nilpotency index $s$. Denote $R_n=R[x]/\langle x^n-1\rangle$. Review the conventional row circulant definition of the QC code. Let $T$ be a cyclic shift operator $T$: $R^N\rightarrow R^N$, which transforms $v=(v_0, v_1, \ldots, v_{N-1})$ into $vT=(v_{N-1}, v_0, \ldots, v_{N-2})$. A linear code $\mathscr{C}$ is called quasi-cyclic (QC) code if it is invariant under $T^\ell$ for some positive integer $\ell$. The smallest $\ell$ such that $T^\ell(\mathscr{C})=\mathscr{C}$ is called the index of $\mathscr{C}$. Clearly, $\ell$ is a divisor of $N$. Let $N=n\ell$. Define a one-to-one correspondence $$\rho: R^N\rightarrow R_n^l$$ $$(a_{0,0}, a_{0,1}, \ldots, a_{0,\ell-1},a_{1,0}, a_{1,1}, \ldots, a_{1,\ell-1}, \ldots, a_{n-1,0}, a_{n-1,1}, \ldots, a_{n-1,\ell-1})$$ $$\mapsto  \underline{a}(x)=(a_0(x), a_1(x), \ldots,  a_{\ell-1}(x))$$ where $a_j(x)=\sum_{i=0}^{n-1}a_{ij}x^i$ for $j=0,1,\ldots,\ell-1$. Then $\mathscr{C}$ is equivalent to for any $\underline{a}(x)=(a_0(x), a_1(x), \ldots, a_{l-1})\in \rho(\mathscr{C})$, $x\underline{a}(x)\in \rho(\mathscr{C})$. Therefore, $\mathscr{C}$ is a QC code if and only if $\rho(\mathscr{C})$ is an $R_n$-submodule of $R_n^\ell$.

\vskip 3mm \par Let $\mathscr{C}=R_n\underline{a}(x)$, where $\underline{a}(x)$ is defined as above. Then $\mathscr{C}$ is called a $1$-generator QC code. Define the annihilator of $\mathscr{C}$ as follows $${\rm Ann}(\mathscr{C})=\{c(x)\in R_n\ |~ c(x)a_i(x)=0, \forall  i=0,1,\ldots, \ell-1\}.$$ It is easy to check that ${\rm Ann}(\mathscr{C})$ is an ideal of $R_n$. Consider the map $\varphi$ from $R_n$ to $\mathscr{C}$, which sends $c(x)\in R_n$ to $c(x)\underline{a}(x)\in \mathscr{C}$. Clearly, $\varphi$ is a surjective $R[x]$-module homomorphism and the kernel of $\varphi$ is ${\rm Ann}(\mathscr{C})$. Therefore $R_n/{\rm Ann}(\mathscr{C})$ is isomorphic to $\mathscr{C}$ and hence $$|\mathscr{C}|=|R_n|/|{\rm Ann}(\mathscr{C})|.$$

\vskip 3mm \par Let $I=\langle a_0(x), a_1(x), \ldots, a_{\ell-1}(x)\rangle$ be the ideal generated by elements $a_0(x)$, $a_1(x)$, $\ldots, a_{\ell-1}(x)\in R_n$. Then, by Theorem 2.2, there exists a unique set of pairwise coprime monic polynomials $F_0, F_1, \ldots, F_s $ over $R$ (possibly, some of them are equal to $1$) such that $F_0F_1\cdots F_s=x^n-1$ and $I=\langle\widehat{F}_1, \gamma\widehat{F}_2,\ldots, \gamma^{s-1}\widehat{F}_s\rangle$, where $\widehat{F}_i=(x^n-1)/F_i$, $i=0,1,\ldots,s$.

\vskip 3mm \noindent {\bf Theorem 5.1 } \emph{Let $\underline{a}(x)=(a_0(x), a_1(x), \ldots, a_{\ell-1}(x))\in R_n^l$, and $\mathscr{C}=R_n\underline{a}(x)$ be a $1$-generator QC code. Let $F_0, F_1, \ldots, F_s$ be given above. Then the annihilator of $\mathscr{C}$ ${\rm Ann}(\mathscr{C})=\langle \widehat{F}_0, \gamma\widehat{F}_s, \ldots, \gamma^{s-1}\widehat{F}_2\rangle$, where $\widehat{F}_i=(x^n-1)/F_i$,
$i=0,1,\ldots,s$.}
\vskip 3mm \noindent\emph{ Proof} Let $I=\langle a_0(x), a_1(x), a_{\ell-1}(x)\rangle$ and ${\rm Ann}(I)$ be the annihilator of $I$. It is easy to verify that ${\rm Ann}(I)={\rm Ann}(\mathscr{C})$. Clearly, $\widehat{F}_0$, $\gamma\widehat{F}_s, \ldots, \gamma^{s-1}\widehat{F}_2\in {\rm Ann}(I)$. On the other hand, since $I\cong R_n/{\rm Ann}(I)$, $$|{\rm Ann}(I)|=|R_n|/|I|=q^{\sum_{i=1}^si{\rm deg}F_{i+1}}=|\langle \widehat{F}_0, \gamma\widehat{F}_s, \ldots, \gamma^{s-1}\widehat{F}_2\rangle|,$$ where $F_{s+1}=F_0$. Therefore ${\rm Ann}(\mathscr{C})={\rm Ann}(I)=\langle \widehat{F}_0, \gamma\widehat{F}_s, \ldots, \gamma^{s-1}\widehat{F}_2\rangle$.  \hfill $\Box$

\vskip 3mm \par Let $F_0, F_1, \ldots, F_s$ be given above. If $F_2=F_3=\cdots=F_s=1$, then $x^n-1=F_0F_1$. Let $\mathscr{C}$ be a $1$-generator QC code generated by $(a_0(x), a_1(x), \ldots, a_{\ell-1}(x))$. Then $I=\langle a _0(x), a_1(x), \ldots, a_{\ell-1}(x))\rangle=\langle F_0\rangle$ and ${\rm Ann}(\mathscr{C})=\langle F_1\rangle$. By Proposition 2.4, we have $I$ and ${\rm Ann}(\mathscr{C})$ are all free cyclic codes over $R$. In the following, we give a sufficient condition for the $1$-generator QC code to be free and a BCH-type bound for the free $1$-generator QC code.
\vskip 3mm \noindent {\bf Theorem 5.2} \emph{Let $\mathscr{C}$ be a $1$-generator QC code with annihilator $\langle F_1\rangle$ generated by $\underline{a}(x)=(a_0(x), a_1(x), \ldots, a_{\ell-1}(x))\in R_n^\ell$.
\vskip 1mm \noindent $(1)$~ $\mathscr{C}$ is a free $1$-generator QC code. Furthermore if ${\rm deg}F_1=k$, then $\mathscr{C}$ is generated by the set $\{{\underline{a}(x), x\underline{a}(x), \ldots, x^{k-1}\underline{a}(x)}\}$ with rank $k$.
 \vskip 1mm \noindent $(2)$ Let $b_i(x)=a_i(x)/F_0$ and ${\rm gcd}(b_i(x), F_1)=1$ for all $i=0,1,\ldots,\ell-1$. Let $\xi^b, \xi^{b+1}, \ldots, \xi^{b+\delta-2}$ be the consecutive roots of $F_0$, where $\xi$ is an $n$-th primitive root of unity in some Galois extension ring ${\mathcal R}$ of $R$. Then the minimum Hamming distance of $\mathscr{C}$ is $d(\mathscr{C})\geq \ell\delta$.}

\vskip 3mm \noindent\emph{ Proof} $(1)$~Define a map as follows $$\phi:~ R[x]\rightarrow R_n^\ell$$ $$ f(x)\mapsto f(x)\underline{a}(x).$$ $\phi$ is an $R[x]$-module homomorphism onto $\mathscr{C}$. Clearly, the kernel of $\phi$ is the annihilator of $\mathscr{C}$. It follows that $\mathscr{C}$ is isomorphic to $R[x]/\langle F_1\rangle$. Since $F_0F_1=x^n-1$, by Theorem 2.1 (5), we have $\mathscr{C}$ is isomorphic to the cyclic code $\langle F_0\rangle$, which is free over $R$. Therefore $\mathscr{C}$ is a free $1$-generator QC code over $R$. If ${\rm deg}F_1=k$, then $\mathscr{C}$ can be generated by the set $\{{\underline{a}(x), x\underline{a}(x), \ldots, x^{k-1}\underline{a}(x)}\}$. From the isomorphism between $\mathscr{C}$ and the free cyclic code $\langle F_0\rangle$, one can deduce the elements in the set are linearly independent over $R$, which implies that the rank of $\mathscr{C}$ is $k$.
\vskip 1mm \noindent $(2)$~For a fixed $i$, $i=0,1,\ldots, \ell-1$,  define a map
$\phi_i$ from $(R[x]/\langle x^n-1\rangle)^\ell$ to $R[x]/\langle x^n-1\rangle$ such that $\phi_i(a_0(x), a_1(x), \ldots, a_{\ell-1}(x))=a_i(x)$. Then $\phi_i(\mathscr{C})$ is a free cyclic code of length $n$ generated by $a_i(x)=b_i(x)F_0$ for all $i=0,1,\ldots, \ell-1$. Since $x^n-1=F_0F_1$, we have one of the components becomes zero if and only if all the others become zero because $b_i(x)F_0p(x)=0$ if and only if $(x^n-1)|b_i(x)F_0p(x)$, i.e., $F_0F_1|b_i(x)F_0p(x)$ if and only if $F_1|b_i(x)p(x)$. Since ${\rm gcd}(b_i(x), F_1)=1$, we have $F_1|p(x)$. Therefore $b_i(x)F_0p(x)=0$ for all $i=0,1,\ldots, \ell-1$. And hence if $c$ is a non-zero codeword in $\mathscr{C}$, then $\phi_i(c)$ is also non-zero. Since ${\rm gcd}(b_i(x), F_1)=1$, it follows that $\phi_i(\mathscr{C})$ is generated by $F_0$. By Proposition 2.5, we have $d(\phi_i(\mathscr{C}))\geq \delta$ implying $d(\mathscr{C})\geq \ell\delta$.                                      \hfill $\Box$

\vskip 3mm \noindent{\bf Example 5.3} Let $R=\mathbb{F}_5+u\mathbb{F}_5$ and $n=24$. Let $\mathscr{C}$ be a $1$-generator QC code with generator $(a_0(x), a_1(x))$ over $R$. Assume that
$a_0(x)=b_0(x)F_0=1\times F_0=x^{16}+2x^{14}+4x^{13}+4x^{12}+3x^{11}+2x^{10}+3x^9+3x^8+3x^7+x^5+2x^4+2x+2$
is a monic divisor of $x^{24}-1$ and $a_1(x)=b_1(x)F_0$, where $b_1(x)=3x^6+2x^5+x^3+2x^2$. Then the annihilator of $\mathscr{C}$ is $\langle F_1\rangle=\langle x^8+3x^6+x^5+3x^3+2x^2+3x+2\rangle$ over $R$. Since $F_0$ having $\xi^i$, $7\leq i \leq18$, among its zeros, where $\xi$ is a $24$-th primitive root of unity in ${\mathcal R}=\mathbb{F}_{5^2}+u\mathbb{F}_{5^2}$, and ${\rm gcd}(b_j(x), F_1)=1$ for $j=0,1$, the cyclic code of length $24$ generated by $a_0(x)$ and $a_1(x)$ is free and the minimum Hamming distance at least $13$. Therefore $\mathscr{C}$ is free with the minimum Hamming distance at least $2\times 13=26$. In fact, the minimum Hamming distance of $\mathscr{C}$ is $30$. Therefore $\mathscr{C}$ is a free $1$-generator QC code of length $48$, with the rank $8$ and the minimum Hamming distance $30$, i.e. an $[48,8,30]$ linear code over $R$.

\vskip 3mm \noindent {\bf 6 Enumeration of $1$-generator quasi-cyclic codes} \vskip 6mm \noindent
In this section, we assume that ${\rm gcd}(n,q)=1$ and ${\rm gcd}(|q|_n, \ell)=1$, where $|q|_n$ denotes the order of $q$ modulo $n$, i.e., the smallest integer $t$ such that $n\mid(q^t-1)$. For simplicity, we denote  $R_n=R[x]/\langle x^n-1\rangle$ and $\langle r\rangle_{\mathscr{R}}$ be the ideal of $\mathscr{R}$ generated by $r\in \mathscr{R}$.

\vskip 3mm \par Let $I=\langle a_0(x), a_1(x), \ldots, a_{\ell-1}(x)\rangle$ be an ideal generated by $a_0(x), a_1(x), \ldots, a_{\ell-1}(x)$ of $R_n$. Then there exist pairwise coprime monic polynomials $F_0, F_1, \ldots, F_s\in R_n$ such that $F_0F_1\cdots F_s=x^n-1$ and $I=\langle\widehat{F}_1, \gamma\widehat{F}_2, \ldots, \gamma^{s-1}\widehat{F}_s\rangle$, where $\widehat{F}_i=(x^n-1)/F_i$, $i=0,1,\ldots,s$. By reduction modulo $\gamma$, we have $\overline{I}=\langle\overline{\widehat{F}}_1\rangle_{\overline{R}_n}$ and
\begin{equation}
{\rm gcd}(\overline{a}_0(x), \overline{a}_1(x), \ldots, \overline{a}_{\ell-1}(x), x^n-1)={\rm gcd}(\overline{\widehat{F}}_1, x^n-1)=\overline{\widehat{F}}_1
\end{equation}
It means that $\overline{F}_1\overline{a}_i(x)=0$ in $\overline{R}_n[x]$, $i=0,1,\ldots,\ell-1$. Assume that there exist polynomials $b_{i1}(x)\in R_n$ such that $$\gamma b_{i1}(x)=F_1a_i(x), ~i=0,1,\ldots,\ell-1. $$ Then $\langle F_1a_0(x), \ldots, F_1a_{\ell-1}(x)\rangle=\langle \gamma b_{01}(x), \ldots, \gamma b_{\ell-1,1}(x)\rangle=\langle \gamma F_1 \widehat{F}_2, \ldots,\gamma^{s-1}F_1\widehat{F}_s\rangle $, which implies that $\langle\overline{b}_{01}(x), \overline{b}_{11}(x), \overline{b}_{\ell-1, 1}(x)\rangle_{\overline{R}_n}=\langle\overline{F}_1\overline{\widehat{F}}_2\rangle_{\overline{R}_n}$ and

\begin{equation}
{\rm gcd}(\overline{b}_{01}(x), \overline{b}_{11}(x), \ldots, \overline{b}_{\ell-1,1}(x), x^n-1)={\rm gcd}(\overline{F}_1\overline{\widehat{F}}_2, x^n-1)=\overline{\widehat{F}}_2
\end{equation}

Repeating the above process, we get the following $s$ equations
\begin{equation} \label{eq:1}
 \begin{cases}
         {\rm gcd}(\overline{a}_0(x), \ldots, \overline{a}_{l-1}(x), x^n-1)={\rm gcd}(\overline{\widehat{F}}_1, x^n-1)=\overline{\widehat{F}}_1 \\
                  {\rm gcd}(\overline{b}_{01}(x),  \ldots, \overline{b}_{\ell-1,1}(x), x^n-1)={\rm gcd}(\overline{F}_1\overline{\widehat{F}}_2, x^n-1)=\overline{\widehat{F}}_2\\
                  \quad  \quad  \quad  \quad  \quad  \quad  \quad  \quad  \quad  \quad  \quad \vdots\\
         {\rm gcd}(\overline{b}_{0,s-1}(x),  \ldots, \overline{b}_{\ell-1,s-1}(x), x^n-1)={\rm gcd}(\overline{F}_1\cdots \overline{F}_{s-1}\overline{\widehat{F}}_s, x^n-1)=\overline{\widehat{F}}_s.
                          \end{cases}
                          \end{equation}

Note that $ \widehat{F}_1, \widehat{F}_2, \ldots, \widehat{F}_s$ are the Hensel lift of $\overline{\widehat{F}}_1, \overline{\widehat{F}}_2, \ldots, \overline{\widehat{F}}_s$, respectively. Therefore we can determine the generators of $I$, i.e., $\widehat{F}_1, \gamma\widehat{F}_2, \ldots, \gamma^{s-1}\widehat{F}_s$. Moreover the $1$-generator QC code $\mathscr{C}$ has the annihilator ${\rm Ann}(\mathscr{C})=\langle \widehat{F}_0, \gamma \widehat{F}_s, \ldots, \gamma^{s-1}\widehat{F}_2\rangle$ if and only if the equations $(7)$ hold.

\vskip 3mm \noindent {\bf Theorem 6.1} \textit{Let $\underline{a}(x)=(a_0(x), a_1(x), \ldots, a_{\ell-1}(x))\in R_n^\ell$ and $\mathscr{C}=R_n\underline{a}(x)$ be a $1$-generator QC code over $R$. Let $\underline{c}(x)=(c_0(x), c_1(x), \ldots, c_{\ell-1}(x))\in R_n^\ell$. Then $R_n\underline{a}(x)=R_n\underline{c}(x)$ if and only if there exists a polynomial $h(x)\in R[x]$ such that $\underline{c}(x)=h(x)\underline{a}(x)$ and $h(x)$ are coprime with the generator of ${\rm Ann}(\mathscr{C})$,i.e. $F=\widehat{F}_0+\gamma \widehat{F}_s+\cdots+\gamma^{s-1}\widehat{F}_2$.}
\vskip 3mm \noindent\emph{ Proof} Since ${\rm Ann}(\mathscr{C})=\langle \widehat{F}_0, \gamma \widehat{F}_s, \ldots, \gamma^{s-1}\widehat{F}_2\rangle$ is an ideal of $R_n$, from the proof process of Theorem 3.6 in [9], we have ${\rm Ann}(\mathscr{C})=\langle \widehat{F}_0+\gamma \widehat{F}_s+\cdots+\gamma^{s-1}\widehat{F}_2\rangle$. Let $\underline{c}(x)=h(x)\underline{a}(x)$ and $h(x)$ and $F$ are coprime. Then $R_n\underline{c}(x)\subseteq R_n\underline{a}(x)$. Since $h(x)$ and $F$ coprime, there exist $u(x), v(x)\in R[x]$ such that $u(x)h(x)+v(x)F=1$. It means that
$R_n\underline{a}(x)=R_n\underline{a}(x)(u(x)h(x)+v(x)F)=R_n\underline{a}(x)u(x)h(x)=u(x)R_n\underline{b}(x)$ in $R_n$ implying $R_n\underline{a}(x)=R_n\underline{c}(x)$
\vskip 1mm  Conversely, if $R_n\underline{a}(x)=R_n\underline{c}(x)$, then there are polynomials $u(x), h(x)\in R[x]$ such that $\underline{c}(x)=h(x)\underline{a}(x)$ and $\underline{a}(x)=u(x)\underline{c}(x)$. Therefore $R_n\underline{a}(x)=u(x)h(x)R_n\underline{c}(x)$, which implies that $(1-u(x)h(x))R_n\underline{a}(x)=0$ in $R_n$. Thus $1-u(x)h(x)\in {\rm Ann}(\mathscr{C})$, i.e. there exists a polynomial $v(x)\in R[x]$ such that $1-u(x)h(x)=v(x)F$. Hence $h(x)$ and $F$ are coprime.                     \hfill $\Box$

\vskip 3mm \par Let $f(x)$ be a basic irreducible polynomial with degree $\ell$ over $R$. Then $\mathcal {R}=R[x]/\langle f(x)\rangle$ be the $\ell$-th Galois extension ring of $R$. We can map $R_n^\ell$ into $\mathcal{R}_n$ via the natural mapping $$\sigma: R_n^\ell \rightarrow \mathcal{R}_n$$ $$(a_0(x), a_1(x), \ldots, a_{\ell-1}(x))\mapsto\sum_{i=0}^{\ell-1}a_i(x)\alpha_i=A(x),$$ where the set $\{\alpha_0, \alpha_1, \ldots, \alpha_{\ell-1}\}$ is a basis for $\mathcal {R}$ over $R$.

\vskip 3mm \par Recall that $I=\langle a_0(x), a_1(x), \ldots,
a_{\ell-1}(x)\rangle=\langle\widehat{F}_1,\gamma\widehat{F}_2, \ldots, \gamma^{s-1}\widehat{F}_s\rangle$. Let $a_i(x)=a_{i1}(x)\widehat{F}_1+\gamma a_{i2}(x)\widehat{F}_2+\cdots +\gamma^{s-1}a_{is}\widehat{F}_s$, where $a_{i1}(x), a_{i2}(x), \ldots, a_{is}(x) \in R[x]$ for each $i=0,1,\ldots,\ell-1$. Then $A(x)=\widehat{F}_1\sum_{i=0}^{\ell-1}a_{i1}(x)\alpha_i+\gamma\widehat{F}_2\sum_{i=0}^{l-1}a_{i2}(x)\alpha_i +\cdots+\gamma^{s-1}\widehat{F}_s \sum_{i=0}^{\ell-1}a_{is}(x)\alpha_i\in \langle\widehat{F}_1, \gamma\widehat{F}_2, \ldots, \gamma^{s-1}\widehat{F}_s\rangle_{\mathcal {R}_n}$.

\vskip 3mm \par Since ${\rm gcd} (n,q)=1$ and ${\rm gcd} (|q|_n, \ell)=1$, the polynomial $x^n-1$ has the same prime factorization in $\mathbb{F}_q[x]$ and$\mathbb{F}_{q^\ell}[x]$. Therefore $x^n-1$ has the same factorization as a product of basic irreducible polynomials over finite chain rings $R$ and $\mathcal{R}$, respectively. In this case, if we assume $\underline{a}(x)=(a_0(x), a_1(x), \ldots, a_{\ell-1}(x))\in R_n^\ell$,
$\overline{A}(x)=\sum_{i=0}^{\ell-1}\overline{a}_i(x)\overline{\alpha}_i$, then ${\rm gcd}(\overline{a}_0(x), \overline{a}_1(x), \ldots, \overline{a}_{\ell-1}(x), x^n-1)= {\rm gcd}(\overline{A}(x),x^n-1)$ ( cf. Lemma 3 in [21]).

 \vskip 3mm \par For all $i=1,2,\ldots,s$, $\widehat{F}_i$ and $F_i$ are coprime in $\mathcal {R}[x]$ because of ${\rm gcd}(n,q)=1$. Therefore there exist $u_i(x), v_i(x)\in \mathcal{R}[x]$ such that $u_i(x)\widehat{F}_1+v_i(x)F_1=1$. Let $e_i=1-v_i(x)F_i=u_i(x)\widehat{F}_i$. Then, by Theorem 2.1 (3), each $e_i$ is the identity of the ring $\langle\widehat{F}_i\rangle_{\mathcal{R}_n}$, $i=1,2,\ldots,s$. If we assume that $e$ is the identity of $\langle\widehat{F}_1, \gamma\widehat{F}_2, \ldots, \gamma^{s-1}\widehat{F}_s\rangle_{\mathcal {R}_n}$, then we have $e=e_1+e_2+\cdots +e_s$, which implies the following direct sum decomposition $$\langle\widehat{F}_1, \gamma\widehat{F}_2, \ldots, \gamma^{s-1}\widehat{F}_s\rangle_{\mathcal {R}_n}=\langle\widehat{F}_1\rangle_{\mathcal {R}_n} \oplus \langle \gamma\widehat{F}_2\rangle_{\mathcal {R}_n}\oplus \cdots \oplus \langle \gamma ^{s-1}\widehat{F}_s\rangle_{\mathcal {R}_n}$$ $$A(x)=A_1(x)+\gamma A_2(x)+\cdots+\gamma^{s-1}A_{s}(x),$$ where $A(x)=\sum_{i=0}^{\ell-1} a_i(x)\alpha_i$, $\gamma ^{j-1}A_j=A(x)e_j$ for all $j=1,2,\ldots,s$.

 \vskip 3mm \noindent {\bf Lemma 6.2}\textit{ Let $A(x)=\sum_{i=0}^{\ell-1}a_i(x)\alpha_i=A_1(x)+\gamma A_2(x)+\cdots+\gamma^{s-1}A_s(x)$, where $a_i(x)\in R_n$ and $\gamma^{j-1}A_{j}(x)\in \langle \gamma^{j-1} \widehat{F}_{j}\rangle_{\mathcal {R}_n}$, $i=0,1,\ldots, \ell-1$, $j=1,2,\ldots,s$. Let $\underline{a}(x)=(a_0(x), a_1(x), \ldots,a_{\ell-1}(x))$. Then $\mathscr{C}=R_n\underline{a}(x)$ is a $1$-generator QC code with annihilator $\langle \widehat{F}_0, \gamma\widehat{F}_s, \ldots, \gamma^{s-1}\widehat{F}_s\rangle$ if and only if $A_j(x)$ and $F_j$ are coprime for all $j=1,2,\ldots,s$.}

\vskip 3mm \noindent\emph{ Proof} Firstly, we will prove that ${\rm gcd}(\overline{A}_j(x), x^n-1)=\overline{\widehat{F}}_j$ if and only if $A_j(x)$ and $F_j$ are coprime for all $j=1,2,\ldots,s$. Let $A_j(x)$ and $F_j$ be coprime over $\mathcal {R}$, $j=1,2,\ldots,s$. Then ${\rm gcd}(\overline{A}_j(x),\overline{F}_j)=1$. On the other hand, since $\gamma^{j-1}A_j(x)\in\langle\gamma^{j-1}\widehat{F}_j\rangle_{\mathcal {R}_n}$, we have $\overline{\widehat{F}}_j \mid{\rm gcd} (\overline{A}_j(x), x^n-1)$, which implies that ${\rm gcd} (\overline{A}_j(x),
x^n-1)=\overline{\widehat{F}}_j$. Conversely, for any $\gamma^{j-1}A_j(x)\in \langle\gamma^{j-1}\widehat{F}_j\rangle_{\mathcal {R}_n}$, if ${\rm gcd} (\overline{A}_j(x), x^n-1)=\overline{\widehat{F}}_j$, then ${\rm gcd} (\overline{A}_j(x), \overline{F}_j)=1$. Therefore,
$A_j(x)$ and $F_j$ are coprime in $\mathcal {R}[x]$.

\par Next we will prove that $\mathscr{C}$ is a $1$-generator QC code generated by $\underline{a}(x)$ with annihilator $\langle \widehat{F}_0, \gamma\widehat{F}_2, \ldots, \gamma^{s-1}\widehat{F}_s\rangle$ if and only if ${\rm gcd}(\overline{A}_j(x), x^n-1)=\overline{\widehat{F}}_j$ for all $j=1,2,\ldots,s$. Note that $$\overline{A}_1(x)=\sum_{i=0}^{\ell-1}\overline{a}_i(x)\overline{\alpha}_i$$ $$\overline{F}_1\overline{A}_2(x)=\sum_{i=0}^{\ell-1}\overline{b}_{1i}(x)\overline{\alpha}_i$$ $$\vdots$$ $$\overline{F}_1\cdots \overline{F}_{s-1}\overline{A}_s(x)=\sum_{i=0}^{\ell-1}\overline{b}_{s-1,i}(x)\overline{\alpha}_i.$$  Then by $${\rm gcd}(\overline{a}_0(x), \overline{a}_1(x), \ldots, \overline{a}_{\ell-1}(x), x^n-1)={\rm gcd}(\overline{A}_1(x), x^n-1)$$ and $${\rm gcd}(\overline{b}_{k0}(x), \overline{b}_{k1}(x), \ldots, \overline{b}_{k,\ell-1}(x), x^n-1)={\rm gcd}(\overline{F}_1\cdots \overline{F}_{k}\overline{A}_{k+1}(x), x^n-1)$$ for all $k=1,2,\ldots,s-1$, we have $\mathscr{C}$ is a $1$-generator QC code with annihilator ${\rm Ann}(\mathscr{C})=\langle \widehat{F}_0, \gamma \widehat{F}_s, \ldots, \gamma^{s-1}\widehat{F}_2\rangle$ if and only if $$\overline{\widehat{F}}_1={\rm gcd}(\overline{a}_0(x), \overline{a}_1(x), \ldots, \overline{a}_{\ell-1}(x), x^n-1)={\rm gcd}(\overline{A}_1(x), x^n-1)$$  $$\overline{\widehat{F}}_{k+1}={\rm gcd}(\overline{b}_{k0}(x), \overline{b}_{k1}(x), \ldots, \overline{b}_{k,\ell-1}(x), x^n-1)={\rm gcd}(\overline{F}_1\cdots \overline{F}_{k}\overline{A}_{k+1}(x), x^n-1)$$ for all $k=1,2,\ldots,s-1$.    \hfill $\Box$

 \vskip 3mm For each $j=1,2,\ldots,s$, define a map $\mu_j$ from finite chain ring $R$ with maximal ideal $\langle \gamma\rangle$ and nilpotency index $s$ to the finite chain ring $R({\rm mod} \gamma^j )=R^{(j)}$ with maximal ideal $\langle \gamma\rangle$ and nilpotency index $j$, where $j$ is a positive integer and $1\leq j\leq s$. Clearly, $\mu_1(R)$ is the residue field $\mathbb{F}_q$ and $\mu_s(R)$ is $R$. Extend the map $\mu_j$ to the polynomial ring $R[x]$, such that for any polynomial $f(x)\in R[x]$, $\mu_j(f(x))=f(x)({\rm mod} \gamma^j )$. Then $\mu_j$ is a surjective ring homomorphism. Therefore we have the following ring isomorphism. $$\phi_j:~ \langle \gamma^{s-j}\widehat{F}_{s-j+1}\rangle_{R_n} \rightarrow \gamma^{s-j}\langle \mu_j(\widehat{F}_{s-j+1})\rangle_{R^{(j)}_n}$$ $$\gamma^{s-j}f(x)\mapsto \gamma^{s-j}\mu_j(f(x)).$$ Denote $\mathscr{G}_j=(\langle \mu_j(\widehat{F}_{s-j+1})\rangle_{{\mathcal R}^{(j)}_n})^*$, the group of units of the ring $\langle \mu_j(\widehat{F}_{s-j+1})\rangle_{{\mathcal R}^{(j)}_n}$.

 \vskip 3mm \noindent {\bf Theorem 6.3} \textit{Let $A(x)=\sum_{i=0}^{\ell-1}a_i(x)\alpha_i=A_1(x)+\gamma A_2(x)+\cdots+\gamma^{s-1}A_s(x)$, where $a_i(x)\in R_n$ and $\gamma^{j-1}A_j(x)\in \langle \gamma^{j-1}\widehat{F}_j\rangle_{{\mathcal R}_n}$ for all $j=1,2,\ldots,s$. Let $\underline{a}(x)=(a_0(x), a_2(x), \ldots,a_{\ell-1}(x))$. Then $\mathscr{C}=R_n\underline{a}(x)$ is a $1$-generator QC code with annihilator $\langle\widehat{F}_0, \gamma\widehat{F}_s, \ldots, \gamma^{s-1}\widehat{F}_2\rangle$ if and only if $\mu_{j}(A_{s-j+1}(x))\in \mathscr{G}_j$ for all $j=1,2,\ldots,s$.}
\vskip 3mm \noindent\emph{ Proof} For $j=1,2,\ldots,s$, if $\mu_{j}(A_{s-j+1}(x))\in \mathscr{G}_j$, then there exists $\mu_j(d(x))\in \mathscr{G}_j$ such that $$\mu_{j}(A_{s-j+1}(x))\mu_j(d(x))=\mu_j(e_{s-j+1})=1-\mu_j(v_{s-j+1}(x))\mu_j(F_{s-j+1}),$$ which implies that $\mu_{j}(A_{s-j+1}(x))\mu_j(d(x))+\mu_j(v_{s-j+1}(x))\mu_j(F_{s-j+1})=1$. Therefore $A_{s-j+1}(x)$ and $F_{s-j+1}$ are coprime for all $j=1,2,\ldots,s$. From Lemma 6.2, $\mathscr{C}$ has the annihilator $\langle\widehat{F}_0, \gamma\widehat{F}_s, \ldots, \gamma^{s-1}\widehat{F}_2\rangle$.
\vskip 1mm Conversely, for each $j=1,2,\ldots,s$, if $A_{s-j+1}(x)$ and $F_{s-j+1}$ are coprime, then there exist polynomials $w_{s-j+1}(x)$, $z_{s-j+1}(x) \in \mathcal
 {R}[x]$ such that $$w_{s-j+1}(x)A_{s-j+1}(x)+z_{s-j+1}(x)F_{s-j+1}=1.$$ It follows that $\mu_j(w_{s-j+1})\mu_j(A_{s-j+1}(x))+\mu_j(z_{s-j+1})\mu_j(F_{s-j+1})=1$ in ring ${\mathcal R}^{(j)}_n$. Multiplying by $\mu_j(e_{s-j+1})=\mu_j(u_{s+j-1}(x)\mu_j(\widehat{F}_{s-j+1})$, we have $$\mu_j(e_{s-j+1})=\mu_j(w_{s-j+1})\mu_j(A_{s-j+1}(x))\mu_j(u_{s+j-1}(x)\mu_j(\widehat{F}_{s-j+1}).$$
 Therefore $\mu_{j}(A_{s-j+1}(x))\in \mathscr{G}_j$ for all $j=1,2,\ldots,s$.           \hfill $\Box$

 \vskip 3mm \noindent {\bf Theorem 6.4} \textit{Let $\mathscr{M}=\{\mathscr{C}|~ \mathscr{C}~is~a~1-generator~QC~code~over~R~with~{\rm Ann}(\mathscr{C})=\langle\widehat{F}_0,\gamma\widehat{F}_s, \ldots, \gamma^{s-1}\widehat{F}_2\rangle\}$. For each $j=1,2,\ldots,s$, let $F_{s-j+1}=f_{s-j+1,1}\ldots f_{s-j+1,r_{s-j+1}}$, where each $f_{s-j+1,p}$ is the monic basic irreducible polynomial with degree $e_{s-j+1,p}$, $p=1,2,\ldots,r_{s-j+1}$. Then $$|\mathscr{M}|=\prod_{j=1}^s\prod_{p=1}^{r_{s-j+1}}\frac{q^{j\ell e_{s-j+1,p}}-q^{(j-1)\ell e_{s-j+1,p}}}{q^{je_{s-j+1, p}}-q^{(j-1)e_{s-j+1,p}}}.$$}

\vskip 3mm \noindent\emph{ Proof} Denote ${\mathcal G}_j=(\langle \mu_j(\widehat{F}_{s-j+1})\rangle_{ R^{(j)}_n})^*$, the group of units of the ring $\langle \mu_j(\widehat{F}_{s-j+1})\rangle_{R^{(j)}_n}$. For simplicity, we write $\mu_j(\widehat{F}_{s-j+1}){\mathcal G}_j\in \mathscr{G}_j/{\mathcal G}_j$ as $\mu_j(\widehat{F}_{s-j+1})$ in this proof process. Define a map as follows
$$\eta:~\mathscr{G}_s/{\mathcal G}_s\times\mathscr{G}_{s-1}/{\mathcal G}_{s-1}\times\cdots\times\mathscr{G}_1/{\mathcal G}_1\rightarrow \mathscr{M} $$
$$(\mu_s(A_1), \mu_{s-1}(A_2), \ldots, \mu_1(A_s))\mapsto R_n\underline{a}(x),$$ where for each $j=1,2,\ldots,s$, $A_j(x)$ is any inverse image of $\mu_{s-j+1}(A_j)$ under the map $\mu_j$, $A(x)=A_1(x)+\gamma A_2(x)+\cdots+\gamma^{s-1}A_s(x)=\sum_{i=0}^{\ell-1}a_i(x)\alpha_i$ and $\underline{a}(x)=(a_0(x), a_1(x), \ldots, a_{l-1}(x))$.
\vskip 1mm For any $\mu_{s-j+1}(A_{1,j}), \mu_{s-j+1}(A_{2,j})\in \mathscr{G}_j/{\mathcal G}_j$, $j=1,2,\ldots,s$, let $$A^{(1)}(x)=\sum_{i=0}^{\ell-1}a_{1i}(x)\alpha_i=A_{11}(x)+\gamma A_{12}(x)+\cdots+\gamma^{s-1}A_{1s}(x),$$
$$A^{(2)}(x)=\sum_{i=0}^{\ell-1}a_{2i}(x)\alpha_i=A_{21}(x)+\gamma A_{22}(x)+\cdots+\gamma^{s-1}A_{2s}(x),$$ $$\underline{a}^{(1)}(x)=(a_{10}(x), a_{11}(x), \ldots, a_{1,\ell-1}(x))$$ and $$\underline{a}^{(2)}(x)=(a_{20}(x), a_{21}(x), \ldots, a_{2,\ell-1}(x)).$$ If $R_n\underline{a}^{(1)}=R_n\underline{a}^{(2)}$, then by Theorem 6.1, there exists a polynomial $h(x)\in R[x]$ such that ${\rm gcd}(h(x), F)=1$. Therefore $$A^{(2)}(x)=\sum_{i=0}^{\ell-1}h(x)a_{1i}(x)\alpha_i=h(x)A_{11}(x)+\gamma h(x)A_{12}(x)+\cdots+\gamma^{s-1}A_{1s}(x).$$ Since for each $j=1,2,\ldots,s$, $e_j$ is the identity of $\langle \widehat{F}_j\rangle_{{\mathcal R}_n}$, we have $\gamma^{j-1}h(x)A_{1j}(x)=\gamma^{j-1}h(x)e_jA_{1j}(x)$. Moreover, $\mu_{s-j+1}(\widehat{F}_j)|\mu_{s-j+1}(h(x))\mu_{s-j+1}(e_j)$ since $\gamma^{j-1}h(x)e_j\in \langle \gamma^{j-1}\widehat{F}_j\rangle$. Then

\begin{equation}
  \begin{array}{l}
    {\rm gcd}(\mu_{s-j+1}(h(x))\mu_{s-j+1}(e_j), x^n-1)\\
    =\mu_{s-j+1}(\widehat{F}_j){\rm gcd}(\mu_{s-j+1}(h(x))\mu_{s-j+1}(e_j), \mu_{s-j+1}(F_j))\\
    =\mu_{s-j+1}(\widehat{F}_j),\\
  \end{array}
\end{equation}
which deduces that ${\rm gcd}(\mu_{s-j+1}(h(x))\mu_{s-j+1}(e_j), \mu_{s-j+1}(F_j))=1$ (the second equality in $(8)$ holds because of $\mu_{s-j+1}(h(x))\mu_{s-j+1}(e_j)$ is a unit in ${\mathcal G}_j$). Using the technique in the proof process of Theorem 6.3, we have $\mu_{s-j+1}(h(x))\mu_{s-j+1}(e_j)\in {\mathcal G}_j$. It follows that $\mu_{s-j+1}(A_{1j}(x))=\mu_{s-j+1}(A_{2j}(x))$ for all $j=1,2,\ldots,s$. Hence $\eta$ is injective.
\vskip 1mm  For any $1$-generator QC code $R_n\underline{a}(x) \in \mathscr{M}$, assume that $A(x)=A_1(x)+\gamma A_2(x)+\cdots+\gamma^{s-1}A_s(x)=\sum_{i=0}^{\ell-1}a_i(x)\alpha_i$, where $\gamma^{j-1}A_j(x) \in \langle \gamma^{j-1}\widehat{F}_j\rangle_{{\mathcal R}_n}$ for all $j=1,2,\ldots,s$. By Theorem 6.3, each $\mu_{j}(A_{s-j+1}(x))\in {\mathcal G}_j$. Moreover $\eta((A_1(x)) , A_2(x),\ldots, A_s(x))=R_n\underline{a}(x)$. Therefore $\eta$ is surjective. It means that $\eta$ is a bijective map.

\vskip 1mm From Theorem 2.1, for each $j=1,2,\ldots,s$, we have $\langle\mu_{j}(\widehat{F}_{s-j+1})\rangle_{{\mathcal R}_n}\cong {\mathcal R}^{(j)}[x]/\langle F_{s-j+1}\rangle$ and $\langle\mu_{j}(\widehat{F}_{s-j+1})\rangle\cong R^{(j)}[x]/\langle F_{s-j+1}\rangle$ implying  $$|\mathscr{M}|=|\bigoplus_{j=1}^s({\mathcal R}^{(j)}[x]/\langle F_{s-j+1}\rangle)^*/(R^{(j)}[x]/\langle F_{s-j+1}\rangle)^*|.$$
Since $F_{s-j+1}=f_{s-j+1,1}\ldots f_{s-j+1, r_{s-j+1}}$, where each $f_{s-j+1,p}$ is the monic basic irreducible polynomial with degree $e_{s-j+1,p}$ over ${\mathcal R}$ and $R$, $p=1,2,\ldots,r_{s-j+1}$, from Theorem 2.1, we have $$|\mathscr{M}|=\prod_{j=1}^s\prod_{p=1}^{r_{s-j+1}}\frac{q^{j\ell e_{s-j+1,p}}-q^{(j-1)\ell e_{s-j+1,p}}}{q^{je_{s-j+1, p}}-q^{(j-1)e_{s-j+1,p}}}.$$                                          \hfill$\Box$

\vskip 3mm \noindent {\bf 7 Generators of $1$-generator quasi-cyclic codes} \vskip 6mm \noindent
In section 6, we have given the enumeration formula of $1$-generator QC codes with the annihilator $\langle \widehat{F}_0,
\gamma\widehat{F}_s, \ldots, \gamma^{s-1}\widehat{F}_2\rangle$. Another problem is that can we find out the one and only one generator of each $1$-generator QC code with the annihilator $\langle \widehat{F}_0,\gamma\widehat{F}_s, \ldots, \gamma^{s-1}\widehat{F}_2\rangle$ over finite chain ring $R$.

 \vskip 3mm \par Let $\underline{a}(x)=(a_0(x), a_1(x), \ldots, a_{\ell-1}(x)) $ and
 $A(x)=\sum_{i=0}^{\ell-1}a_i(x)\alpha_i=A_1(x)+\gamma A_2(x)+\cdots+\gamma^{s-1}A_s(x)$, where the set $\{\alpha_0,\alpha_1,\ldots, \alpha_{\ell-1}\}$ is a basis for $\mathcal {R}$ over $R$. From the proof of Theorem 6.4, if we find $\mu_{j}(A_{s-j+1}(x))$ for all $j=1,2,\ldots,s$, we can determine the generator of $R_n\underline{a}(x)$, i.e., $(a_0(x), a_1(x), \ldots, a_{\ell-1}(x))$.  By Theorem 2.1 $(5)$, we have

 $$({\mathcal R}^{(j)}[x]/\langle F_{s-j+1}\rangle)^*/(R^{(j)}[x]/\langle F_{s-j+1}\rangle)^*\cong \bigoplus_{p=1}^{r_{s-j+1}}({\mathcal R}^{(j)}[x]/\langle f_{s-j+1,p}\rangle)^*/(R^{(j)}[x]/\langle f_{s-j+1,p}\rangle)^*,$$
 where for each $j=1,2,\ldots,s$, $F_{s-j+1}=f_{s-j+1,1}f_{s-j+1,2}\ldots f_{s-j+1,r_{s-j+1}}$ and each $f_{s-j+1,p}$ is a monic basic irreducible polynomial with degree $e_{s-j+1, p}$ over $R$ and ${\mathcal R}$, $p=1,2,\ldots,r_{s-j+1}$. Therefore $$\mu_j(A_{s-j+1}(x))=\mu_j(e_{s-j+1})\sum_{p=1}^{r_{s-j+1}}\varepsilon_{jp},$$ where $e_{s-j+1}$ is the identity of $\langle \widehat{F}_{s-j+1}\rangle_{{\mathcal R}_n}$ and $\varepsilon_{jp}$ is the representation of the quotient group $({\mathcal R}^{(j)}[x]/\langle f_{s-j+1,p}\rangle)^*/(R^{(j)}[x]/\langle f_{s-j+1,p}\rangle)^*$ for each $j=1,2,\ldots,s$ and $p=1,2,\ldots r_{s-j+1}$.

\vskip 3mm For simplicity, we choose the set $\{1, \alpha_1, \ldots, \alpha_{{\ell-1}}\}$ as the basis of ${\mathcal R}$ over $R$. Denote the set $$\mathcal {T}_{s-j+1,p}=\{\beta_{p1}\overline{\alpha}_1+\cdots+\beta_{p,\ell-1}\overline{\alpha}_{\ell-1}|~\beta_{p1}, \ldots, \beta_{p,\ell-1} \in \mathbb{F}_q[x]/\langle f_{s-j+1, p}\rangle\}.$$
For each $p=1,2,\ldots,r_{s-j+1}$, let $\xi_p$ be an element in ${\mathcal R}^{(j)}[x]/\langle f_{s-j+1, p}\rangle$ with order $q^{\ell e_{s-j+1,p}}-1$ and $\mathcal {X}_p=\{0, 1, \ldots, q^{\ell e_{s-j+1,p}}-2, \infty\}$, where we adopt $\xi_p^\infty=0$.

 \vskip 3mm \noindent {\bf Theorem 7.1} \textit{ For $p=1,2,\ldots,r_{s-j+1}$, let the set
 \begin{equation}
  \begin{array}{c}
    \mathcal {Q}^j_p=\{\xi_p^a+\gamma \xi_p^{a+b_1}+\cdots+\gamma^{j-1}\xi_p^{a+b_{j-1}}|~a=0,1,\ldots, \frac{q^{\ell e_{s-j+1,p}}-1}{q^{e_{s-j+1,p}}-1};\\
    b_1,\ldots,b_{s-1}\in\mathcal {X}_p~and~\overline{\xi}^{b_1}, \ldots, \overline{\xi}^{b_{j-1}}\in \mathcal {T}_{s-j+1,p}\}
  \end{array}.
\end{equation}
Then $\mathcal {Q}^j_p$ is the complete set of representation of the cosets of $$({\mathcal R}^{(j)}[x]/\langle f_{s-j+1,p}\rangle)^*/(R^{(j)}[x]/\langle f_{s-j+1,p}\rangle)^*.$$}

 \vskip 3mm \noindent\emph{ Proof}  Clearly, $$|\mathcal {Q}^j_p|=\frac{q^{\ell e_{s-j+1},p}-1}{q^{e_{s-j+1}}-1}\underbrace{\frac{q^{\ell e_{s-j+1,p}}}{q^{e_{s-j+1,p}}}\cdots \frac{q^{\ell e_{s-j+1,p}}}{q^{e_{s-j+1,p}}}}_{j-1}=\frac{q^{j\ell e_{s-j+1,p}}-q^{(j-1)\ell e_{s-j+1,p}}}{q^{je_{s-j+1, p}}-q^{(j-1)e_{s-j+1,p}}},$$ which implies that $$|\mathcal {Q}^j_p|=({\mathcal R}^{(j)}[x]/\langle f_{s-j+1,p}\rangle)^*/(R^{(j)}[x]/\langle f_{s-j+1,p}\rangle)^*.$$ Therefore we only need to prove that different elements in $\mathcal {Q}^j_p$ belong to different cosets. For any $A=\xi_p^{a_1}+\gamma \xi_p^{a_1+b_{11}}+\cdots +\gamma^{j-1}\xi_p^{a_1+b_{j-1,1}}$ and $B=\xi_p^{a_2}+\gamma \xi_p^{a_2+b_{12}}+\cdots +\gamma^{j-1}\xi_p^{a_2+b_{j-1,2}}$ in $({\mathcal R}^{(j)}[x]/\langle f_{s-j+1,p}\rangle)^*/(R^{(j)}[x]/\langle f_{s-j+1,p}\rangle)^*$, if $A=B$ then $AB^{-1}\in (R^{(j)}[x]/\langle f_{s-j+1,p}\rangle)^*$. It follows that $\overline{A}~\overline{B}~^{-1}\in (\mathbb{F}_{q}[x]/\langle f_{s-j+1,p}\rangle)^*$. Therefore $\frac{q^{\ell e_{s-j+1},p}-1}{q^{e_{s-j+1,p}}-1}|(a_1-a_2)$, which deduces $a_1=a_2$. Repeating the above process, we have $b_{\lambda1}=b_{\lambda2}$, where $\lambda=1,2, \ldots, j-1$.                        \hfill$\Box$

\vskip 3mm \noindent {\bf Theorem 7.2} \emph{For each $j=1,2,\ldots,s$, let $F_{s-j+1}=f_{s-j+1,1}\ldots f_{s-j+1,r_{s-j+1}}$, where each $f_{s-j+1,p}$ is the monic basic irreducible polynomial with degree $e_{s-j+1,p}$, $p=1,2,\ldots,r_{s-j+1}$. Let $e_{s+j-1}$ be the identity of $\langle \widehat{F}_{s-j+1}\rangle_{{\mathcal R}_n}$. Then the elements $$\sum_{j=1}^s\sum_{p=1}^{r_{s-j+1}}\gamma^{s-j}\mu_j(e_{s-j+1})\varepsilon_{jp}$$ determine all the distinct $1$-generator QC codes with annihilator $\langle \widehat{F}_0, \gamma \widehat{F}_s, \ldots, \gamma^{s-1}\widehat{F}_2\rangle$.}
\vskip 3mm \noindent\emph{ Proof} Since $A(x)=\sum_{i=0}^{\ell-1} a_i(x)\alpha_i$, it follows that if we get $A(x)$ we can determine the generator of the QC code. For $A(x)$, we have $A(x)=\sum_{j=1}^s\gamma^{j-1}A_j(x)=\sum_{j=1}^s\gamma^{s-j}A_{s-j+1}(x)\cong
\sum_{j=1}^s\gamma^{s-j}\mu_j(A_{s-j+1}(x))=\sum_{j=1}^s\gamma^{s-j}\mu_j(e_{s-j+1})\sum_{p=1}^{r_{s-j+1}}\varepsilon_{jp}
=\sum_{j=1}^s\sum_{p=1}^{r_{s-j+1}}\gamma^{s-j}\mu_j(e_{s-j+1})\varepsilon_{jp}$. It has proved the result.  \hfill$\Box$

\vskip 3mm For all $j=1,2,\ldots,s$, if we let $A_{s-j+1}(x)$ be any inverse image of $\mu_j(A_{s-j+1}(x))$ under the map $\mu_j$ in ${\mathcal R}$, then Theorem 7.2 gives an algorithm of finding the generator $(a_0(x), a_1(x), \ldots, a_{\ell-1}(x))$ over $R$ actually.

\vskip 3mm In the rest of this section, we give an example to illustrate the application of Theorem 6.4 and Theorem 7.2.

\vskip 3mm \noindent {\bf Example 7.3} Let $\ell=2$ and $n=7$. Then $x^7-1$ can be factored into a product of basic irreducible polynomials as $ x^7-1=(x+1)(x^3+x+1)(x^3+x^2+1)$ over $R=\mathbb{F}_2+u\mathbb{F}_2$. Let $F_0=x^3+x+1$, $F_1=x+1$ and $F_2=x^3+x^2+1$. Consider the $1$-generator QC code $\mathscr{C}$ with annihilator $\langle \widehat{F}_0, u\widehat{F}_2\rangle$. Therefore in language of Theorem 6.4, we have $s=2$, $\ell=2$, $e_{21}=3$, $e_{11}=1$. By Theorem 6.4, we have
\begin{equation}
  \begin{array}{cl}
    |\mathcal {M}| & =\prod_{j=1}^2\frac{2^{j\times 2 \times e_{3-j}}-2^{(j-1)\times 2\times e_{3-j}}}{2^{j \times e_{3-j}}-2^{(j-1)\times e_{3-j}}}\\
                   & =\frac{2^{1\times 2\times 3}-1}{2^{1\times 3}-1}\times \frac{2^{2\times 2\times 1}-2^{1\times 2\times1}}{2^{2\times 1}-2^{1\times 1}}\\
                   & =9\times 6   \\
                   & =54.            \\
  \end{array}
\end{equation}

\par Let ${\mathcal R}^{(2)}=R^{(2)}[x]/\langle x^2+x+1\rangle$ and $\xi=x+\langle x^2+x+1\rangle$. Then $\xi$ is an element of order $2^2-1=3$ in ${\mathcal R}^{(2)}$. Choose a basis of ${\mathcal R}$ over $R$ as $\{ 1, \xi\}$. Since $F_1=x-1$, we have ${\mathcal R}^{(2)}[x]/\langle F_1\rangle={\mathcal R}^{(2)}$. Then $\mu_2(e_1)=\widehat{F}_1=x^6+x^5+x^4+x^3+x^2+x+1$ is the identity of $\langle \widehat{F}_1\rangle_{{\mathcal R}^{(2)}_n}$, and $\xi_1=\xi F_0=\xi x^6+\xi x^5+\xi x^4+\xi x^3+\xi x^2+\xi x+\xi$ is an element of order $2^2-1=3$ in ${\mathcal R}^{(2)}[x]/\langle F_1\rangle$. In the language of Theorem 7.1, we have $\mathcal {T}_{11}=\{ \overline{\xi}\,\beta_{11}|\, \beta_{11} \in F_2\}=\{ 0, \overline{\xi}\}$ and $\mathcal {Q}_1^2=\{ \xi_1^a+u\xi_1^{a+b}\, | \, a=0,1,2; \overline{\xi}_1^b \in \mathcal {T}_{11}\}=\{ F_0, \xi F_0, (1+\xi) F_0, (1+u\xi) F_0, (1+u+\xi) F_0, ((1+u)\xi+u) F_0\}$, which is the complete set of representatives of cosets of $({\mathcal R}^{(2)}[x]/\langle F_1\rangle)^*/(R^{(2)}[x]/\langle F_1\rangle)^*$.

\vskip 1mm In the language of Theorem 7.1, we have ${\mathcal R}^{(1)}=F_{2^2}$ and ${\mathcal R}^{(1)}[x]/\langle
x^3+x^2+1\rangle=\mathbb{F}_{2^6}$ with the primitive element $\xi_1=x^5+\overline{\xi}x^4+x^3+\overline{\xi}^2x^2+\overline{\xi}^2x+\overline{\xi}$ and the identity $\mu_1(e_2)=x^4+x^2+x+1$, respectively. Then $\mathcal {Q}_1^1=\{ \xi_1^a|~ a=0,1,\ldots,8\}$. By Theorem 7.2, we get $A(x)=A_1(x)+uA_2(x)\cong \mu_2(e_1)\varepsilon_{21}+u\mu_{1}(e_2)\varepsilon_{11}$, where $\varepsilon_{21}\in \mathcal {Q}_1^2$, $\varepsilon_{11}\in \mathcal {Q}_1^1$ and $A_1(x)$, $A_2(x)$ are any inverse images of $\mu_2(e_1)\varepsilon_{21}$, $\mu_{1}(e_2)\varepsilon_{11}$ under the map $\mu_2$, $\mu_1$, respectively.

\vskip 1mm Clearly, the length of $\mathscr{C}$ is $14$ and $|\mathscr{C}|=\frac{4^7}{4^3 \times 2^3}=4\times 2^3=32$. Define the Lee weight of the elements $0, 1, u, 1+u$ of $\mathbb{F}_2+u\mathbb{F}_2$ as $0, 1, 2, 1$, respectively. Moreover, the Lee weight of an $n$-tuple in $R^n$ is the sum of the Lee weights of its components. The Gray map $\varphi$ sends the elements $0,1,u,1+u$ of $R$ to $(0,0), (0,1), (1,1), (1,0)$ over $\mathbb{F}_2$, respectively. It is easy to verify that $\varphi$ is a linear isometry from $R^n$ (Lee distance) to $\mathbb{F}_2^{2n}$ (Hamming distance). Therefore we have $\varphi(\mathscr{C})$ is a linear code of length $28$ with dimension $5$ over $\mathbb{F}_2$. Let $g_0=1+x+x^2+x^3+x^4+x^5+x^6$, $g_1=1+x+x^2+x^4$, $g_2=1+x^3+x^4+x^5$ and $g_3=1+x^3+x^6$. In the following table, we list the generators and the minimum distances of all distinct $1$-generator QC codes with annihilator $\langle\widehat{ F}_0, u\widehat{F}_2\rangle$ over $R$.

\vskip 3mm \noindent {\bf 8 Conclusion} \vskip 6mm \noindent
In this paper, we mainly consider QC codes over finite chain rings. Module structures and trace representation of QC codes are studied, which lead to some distance bounds on the minimum Hamming distance of QC codes. Particularly, we investigate the $1$-generator QC code giving the explicit structure of its annihilator that be used to calculate the number of codewords in QC code. Moreover, under some conditions, we discuss the enumerator and the generator of the $1$-generator QC code with some fixed annihilator.

\vskip 3mm More recently, there are some research papers on \emph{quasi-twisted} (QT) codes, which are natural generalizations of QC codes (see [1] [8] [11] [15]). They also have good algebra properties and can produce some good linear codes over finite fields [1,8,11]. In [15], the author used the \emph{generalized discrete Fourier transform} (GDFT) to study structural properties of QT codes of arbitrary lengths over finite fields. At the end of [15], the author gave a construction algorithm to construct QT codes. If we assume the block length of the QT code is coprime with the characteristic of the finite chain ring, then all the results in this paper are valid for the QT code. But the structural properties for QT codes of arbitrary lengths over finite chain rings are also interesting open problems for further consideration.

\vskip 3mm \noindent {\bf Acknowledgments}  \emph{This research is supported by the National Key Basic Research Program of China (Grant No. 2013CB834204), and the National Natural Science Foundation of China (Grant Nos. 61171082, 10990011 and 60872025).}

\begin{table}[ht]
\begin{center}
\begin{small}
\begin{tabular}{|c|c|c|}
\hline
Generators  &  Minimum Lee distance &  Minimum Hamming distance\\
\hline
$(g_0+ug_1, 0)$ & $6$ & 3\\
\hline
$(g_0+uxg_1,ug_1)$ & $7$ & 7\\
\hline
$(g_0+g_3,ug_1)$ & $7$ & 7\\
\hline
$(g_0,uxg_2)$ & $7$ & 4\\
\hline
$(g_0+uxg_2,ug_1)$ & $7$ & 7\\
\hline
$(g_0+uxg_2,uxg_3)$ & $7$  & 7\\
\hline
$(g_0+uxg_1,uxg_1)$ & $7$ & 7\\
\hline
$(g_0+uxg_2,ux^2g_1)$ & $7$ & 7\\
\hline
$(g_0+ug_2,ug_1)$  &  $7$ & 7\\
\hline

$(ug_1,g_0)$  & $7$ &4\\
\hline
$(uxg_1,g_0+ug_1)$ & $7$ & 7\\
\hline
$(ug_3,g_0+ug_1)$ & $7$ & 7\\
\hline
$(0,g_0+uxg_2)$ & $6$ & 3\\
\hline
$(uxg_2,g_0+ug_1)$ & $7$ & 7\\
\hline
$(uxg_2,g_0+ug_3)$ & $7$ & 7\\
\hline
$(uxg_1,g_0+uxg_1)$ & $7$ & 7\\
\hline
$(uxg_2,g_0+ux^2g_1)$  & $7$ & 7\\
\hline
$(ug_2,g_0+ug_1)$  &  $7$ & 7\\
\hline

$(g_0+ug_1,g_0)$  & $8$ & 4\\
\hline
$(g_0+uxg_1,g_0+ug_1)$ & $12$ & 6\\
\hline
$(g_0+ug_3,g_0+ug_1)$ & $12$ & 6\\
\hline
$(g_0,g_0+uxg_2)$ & $8$ & 4\\
\hline
$(g_0+uxg_2,g_0+ug_1)$ & $12$ & 6\\
\hline
$(g_0+uxg_2,g_0+ug_3)$ & $12$ & 6\\
\hline
$(g_0+uxg_1,g_0+uxg_1)$ & $12$ & 6\\
\hline
$(g_0+uxg_2,g_0+ux^2g_1)$  & $12$ & 6\\
\hline
$(g_0+ug_2,g_0+ug_1)$  &  $12$ & 6\\
\hline

$(g_0+ug_1, ug_0+ug_0)$  & $6$ & 3\\
\hline
$(g_0+uxg_1,ug_1+ug_0)$ & $13$ & 7\\
\hline
$(g_0+g_3,ug_1+ug_0)$ & $13$ & 7\\
\hline
$(g_0,uxg_2+ug_0)$ & $8$ & 4\\
\hline
$(g_0+uxg_2,ug_1+ug_0)$ & $13$ & 7\\
\hline
$(g_0+uxg_2,uxg_3+ug_0)$ & $13$ & 7\\
\hline
$(g_0+uxg_1,uxg_1+ug_0)$ & $13$ & 7\\
\hline
$(g_0+uxg_2,ux^2g_1+ug_0)$  & $13$ & 7\\
\hline
$(g_0+ug_2,ug_1+ug_0)$  &  $13$ & 7\\
\hline

$(g_0+ug_1+ug_0,g_0)$  & $8$ & 4\\
\hline
$(g_0+uxg_1+ug_0,g_0+ug_1)$ & $12$ & 6\\
\hline
$(g_0+ug_3+ug_0,g_0+ug_1)$ & $12$ & 6\\
\hline
$(g_0+ug_0,g_0+uxg_2)$ & $8$ & 4\\
\hline
$(g_0+uxg_2+ug_0,g_0+ug_1)$ & $12$ & 6\\
\hline
$(g_0+uxg_2+ug_0,g_0+ug_3)$ & $12$ & 6\\
\hline
$(g_0+uxg_1+ug_0,g_0+uxg_1)$ & $12$ & 6\\
\hline
$(g_0+uxg_2+ug_0,g_0+ux^2g_1)$  & $12$ & 6\\
\hline
$(g_0+ug_2+ug_0,g_0+ug_1)$  &  $12$ & 6\\
\hline
\end{tabular}
\end{small}
\end{center}
\end{table}

\begin{table}[ht]
\begin{center}
\begin{small}
\begin{tabular}{|c|c|c|}
\hline
Generators  &  Minimum Lee distance &  Minimum Hamming distance \\
\hline

$(ug_1+ug_0,g_0+ug_0)$  & $8$ & 4\\
\hline
$(uxg_1+ug_0,g_0+ug_1+ug_0)$ & $13$ & 7\\
\hline
$(ug_3+ug_0,g_0+ug_1+ug_0)$ & $13$ & 7\\
\hline
$(ug_0,g_0+uxg_2+ug_0)$ & $6$ & 3\\
\hline
$(uxg_2+ug_0,g_0+ug_1+ug_0)$ & $13$ & 7\\
\hline
$(uxg_2+ug_0,g_0+ug_3+ug_0)$ & $13$ & 7\\
\hline
$(uxg_1+ug_0,g_0+uxg_1+ug_0)$ & $13$ & 7\\
\hline
$(uxg_2+ug_0,g_0+ux^2g_1+ug_0)$  & $13$ & 7\\
\hline
$(ug_2+ug_0,g_0+ug_1+ug_0)$  &  $13$ & 7\\
\hline
$(ug_1+ug_0,g_0+ug_0)$  & $8$ & 4\\
\hline
$(uxg_1+ug_0,g_0+ug_1+ug_0)$ & $13$ & 7\\
\hline
$(ug_3+ug_0,g_0+ug_1+ug_0)$ & $13$ & 7\\
\hline
$(ug_0,g_0+uxg_2+ug_0)$ & $6$ & 3\\
\hline
$(uxg_2+ug_0,g_0+ug_1+ug_0)$ & $13$ & 7\\
\hline
$(uxg_2+ug_0,g_0+ug_3+ug_0)$ & $13$ & 7\\
\hline
$(uxg_1+ug_0,g_0+uxg_1+ug_0)$ & $13$ & 7\\
\hline
$(uxg_2+ug_0,g_0+ux^2g_1+ug_0)$  & $13$ & 7\\
\hline
$(ug_2+ug_0,g_0+ug_1+ug_0)$  &  $13$ & 7\\

\hline
\end{tabular}
\end{small}
\end{center}
\end{table}


\end{document}